\documentclass[letterpaper,12pt]{article}

\usepackage[numbers,sort&compress]{natbib}                   
\usepackage{doi} 
 
 \usepackage{graphicx,hyperref,ulem,xcolor}
 \usepackage{color}
 \usepackage{multirow}
\usepackage{physics}
\usepackage{slashed}
\usepackage[font=small]{caption}
\usepackage{float}
\usepackage{dsfont}
\usepackage{fancyhdr}
 \usepackage{amsmath}
 \usepackage{amsfonts}
 \usepackage{amssymb}
 \usepackage{rotating}
\usepackage{comment}
\usepackage[T1]{fontenc}        
\usepackage[utf8]{inputenc}

\providecommand{\keywords}[1]
{
  \small	
  \textbf{\textit{Keywords---}} #1
}

 \definecolor{cviolet}{rgb}{0.4, 0, 0.6}
 \definecolor{dgreen}{rgb}{0, 0.6, 0.1}
 
 \def\bea{\begin{eqnarray} }
 \def\eea{ \end{eqnarray} }

\newcommand{\sy}{$S_3~$}

\begin{document}

 \title{Quark mixing model with three Higgs doublets and \sy modular symmetry  at $\tau=i$  }
 
\author{ \small{C. Cer\'on and M. Mondrag\'on} \\
  \footnotesize{ 
  Instituto de F\'{\i}sica, Universidad Nacional Autónoma de México,}\\
\footnotesize{ Ciudad de México, C.P. 04510, 
Mexico}}
  
\maketitle 

\begin{abstract}
  We study the quark sector of a model with three Higgs electroweak doublets, complemented with a  modular $S_3$ symmetry. We show that  a proper assignment of the quark and Higgs fields in $S_3$ and their modular weights allows to write a mass matrix with a phenomenologically viable texture.  Moreover, by considering the conditions on the Higgs VEVs derived from the minimization of the scalar potential  together with evaluation of the Yukawa couplings at the fixed point $\tau=i$, leads to a significant reduction of the free parameters.  We present two setups, one with three effective parameters, which is overly constrained, and one with four effective parameters, where the $V_{CKM}$ matrix can be fitted precisely.
  
  \end{abstract}

\keywords{
Standard Model, Mass hierarchy, \sy symmetry, $V_{CKM}$ matrix, Modular symmetry, Modular forms, Yukawa sector }

\section{Introduction}
\label{introduction}
An open question that the Standard Model (SM) has not been able to answer concerns its flavor structure. A common way to address this issue is to add scalars, usually together with right-handed neutrinos, and often supplemented with flavour symmetries. In this context, several models based on discrete groups have been proposed, such as $S_3$, $A_4$, $S_4$, among others, where good results have been obtained (for a review on discrete symmetries in model building see \cite{Ishimori:2010au}). Following this direction, one way to approach the problem is to extend the Standard Model with the aforementioned finite groups but starting from the modular group $\Gamma$, which has been widely used in string theory and other areas \cite{lam20012,dixon1987conformal,Altarelli:2005yx,CHEN2020135153,Feruglio:2017spp,Meloni2023,Penedo:2024gtb,Feruglio:2024ytl,Higaki:2024pql,Jung:2024bgi,Kobayashi:2025hnc}. This group could give an explanation to the flavor structure by means of holomorphic functions known as modular forms \cite{diamond2005first}, although it has been found out recently that it is also possible to have non-holomorphic modular forms \cite{Qu:2024rns}.  Yukawa couplings, and therefore masses and mixing angles,  may be described in terms of modular forms with a particular transformation under the $\Gamma$ group. The above makes the difference between directly using the discrete groups and the modular group, thus providing an apparent advantage for the new restrictions that appear. Since there is no unique assignment of the modular symmetries to the fields, this advantage  might be lost, with the proliferation of extra parameters given by the modular charges of the fields.

Many modular flavour models have already been constructed, with and without supersymmetry \cite{CHEN2022136843,PhysRevD.91.126014,universe9120512,Ratz:2024imd,Okada:2025jjo,Ding2025,Qu2024,NOMURA2019134799,Qu:2025ddz,Li:2024svh}. 
Modular symmetries might also be combined to describe different sectors, for instance in ~\cite{kobayashi2019finite} the authors use modular forms to find mixing angles and mass matrices using modular $A_4$ for lepton masses and modular $S_3$ for the  CKM matrix.
 In \cite{feruglio2019neutrino} modular forms are used in the framework of supersymmetry to describe  neutrino masses by means of the Weinberg operator and the seesaw mechanism. Also, in 
\cite{PhysRevD.98.016004,penedo2019lepton,novichkov2019modular,deMedeirosVarzielas2023}, they use $\Gamma_2$, $\Gamma_3$, $\Gamma_4$ and $\Gamma_5$ which correspond to $S_3$, $A_4$, $S_4$ and $A_5$, respectively, as alternative ways to use the modular group $\Gamma$.

Since the $S_3$  symmetry  model with three Higgs doublets has proven to be successful  in explaining the masses and mixing of quarks and leptons \cite{Kubo:2003iw,canales2013quark,Mondragon:2007af}, special attention will be paid to this symmetry focusing on the advantages that are gained by assuming a modular symmetry over a conventional flavor symmetry. To achieve this, we will consider $\Gamma_2$ whose isomorphism is $S_3$, in addition, three Higgs doublets will be introduced together with their scalar potential invariant under $S_3$. By exploiting the symmetries inherent to \sy as a modular group, and the results in the minimization of the \sy potential with three Higgs doublets \cite{Das:2014fea,Gomez-Bock:2021uyu}, we show that it is possible to describe  the quark mass matrices and the $V_{CKM}$ matrix successfully with few parameters through a two texture zero matrix. In \cite{Petcov20231,Petcov20232} the authors start with a diagonal $V_{CKM}$ matrix evaluated at the modular fixed points, and to obtain the small non-zero entries a small deviation of this symmetry is assumed. In here, we also take advantage of these modular fixed points and the inherent symmetries of $S_3$, but in a different way.  The minimization of the most general scalar potential with an \sy symmetry (non-modular) leaves an unbroken $Z_2$ symmetry \cite{Das:2014fea,Gomez-Bock:2021uyu}, which reflects in an unrealistic block structure for the $V_{CKM}$ matrix. In our case,  by using instead modular \sy and combining this residual symmetry with the evaluation of the Yukawa couplings exactly at one of the modular fixed points, we both reduce drastically the number of free parameters and find a phenomenologically viable structure for the mixing matrix. We show how this way it is possible to extract the values of the Lagrangian parameters in terms of the values obtained from the fit and  one parameter of the scalar potential. 

This work is organized as follows: In section \ref{grupointro}, a short introduction to modular groups and its relationship with some finite groups is made.  We summarize the properties of S3 and its modular version. Section \ref{modelo} contains the application of the finite modular group $\Gamma_2$ to a model of the Yukawa sector, with three Higgs doublets and the corresponding assignments. In section \ref{calculationsmass} we propose two different setups for free parameters to obtain  the texture zero matrix. We also present the elements of the quark mixing matrix, $V_{CKM}$, and the respective comparison with the experimental data by means of a function $\chi^2$. 

\section{ Modular group}
\label{grupointro}
Modular symmetries have already been implemented in some areas of physics such as models of magnetized branes \cite{cremades2004computing,kobayashi2018modularstring}, or compactification of orbifolds \cite{spalinski1992duality,hamidi1987interactions,erler1992dependence,lauer1989duality}, among others \cite{nomura2020modular,nomura2021inverse,nomura2021two,chen2020note,novichkov2021fermion,novichkov2021double}. On the other hand, the symmetry group $S_3$ mentioned in the previous section has been used successfully to address the mixing patterns of both quarks and leptons \cite{Kubo:2003iw,canales2013quark,Mondragon:2007af,harrison2003permutation}.
An interesting alternative is to propose a modular symmetry through the isomorphism between the finite modular group $\Gamma_2$ and $S_3$. In this section a brief description will be made of the definition and properties of the modular group as well as the modular forms and their relationship with $S_3$. Some of the information mentioned here can be found in Refs.~ \cite{feruglio2019neutrino,liu2019neutrino,schoeneberg2012elliptic,gunning2016lectures,diamond2005first,katok1992fuchsian,deAdelhartToorop:2011re}.

In order to define the finite modular groups, the modular group $\Gamma$ must first be defined as 
\begin{equation}
    \Gamma= SL_2(\mathbb{Z})=\left\lbrace \left(
    \begin{array}{cc}      
        a & b \\ 
        c & d
    \end{array}  
    \right)|a,b,c,d \in \mathbb{Z}, ad-bc=1 \right\rbrace.
\end{equation}
From this group, the inhomogeneous (fractional) transformation $\gamma$ over a parameter $\tau$ is defined as
\begin{eqnarray}
    \gamma(\tau) =\left(
    \begin{array}{cc}
        a & b \\ 
        c & d
    \end{array}  
    \right)(\tau)\rightarrow\frac{a\tau +b}{c \tau+d},
\end{eqnarray}
with $\tau \in \hat{\mathbb{C}}$ \hspace{0.3cm} ($\hat{\mathbb{C}}=\mathbb{C}-\{ \infty \}$). It should be noted that the matrices $\mathbf{1}_{2\cross 2}$ and $-\mathbf{1}_{2\cross 2}$ lead to the same transformation, so the matrices associated with $ \gamma$ form the group $\overline{\Gamma}\equiv PSL_2(\mathbb{Z})=SL_2(\mathbb{Z})/\{ \mathbf{1}_{2\cross 2},-\mathbf{1}_{2\cross 2}\}$.
This group can be generated by
\begin{align}
    S: \tau \rightarrow -\frac{1}{\tau}, \notag \\
    T: \tau \rightarrow \tau +1,
\end{align}
which in $\Gamma$ corresponds to the matrices
\begin{equation}
    T=\left(
        \begin{array}{cc}
            1 & 1 \\ 
            0 & 1
        \end{array}  
    \right) \hspace{1cm}\text{and} \hspace{1cm}S=\left(
        \begin{array}{cc}
            0 & 1 \\ 
            -1 & 0
        \end{array}  
    \right),
\end{equation}
they must satisfy
\begin{equation}
    S^2= \mathbf{1} \hspace{1cm} \text{and} \hspace{1cm} (ST)^3= \mathbf{1}.
\end{equation}

An important feature of this transformation is that it applies only to the upper complex half-plane $\mathcal{H}$ defined as
\begin{equation}
    \mathcal{H}=\{\tau \in \mathbb{C}: \mathit{Im}(\tau)>0 \}.
\end{equation}
since $\tau \in \mathcal{H}$, then,  $\mathit{Im}[\tau]>0$. This allows us to conclude that  $\mathit{Im}[\gamma(\tau)]>0$, that is, $\gamma$ acts from $\mathcal{H}$ to $\mathcal{H}$.\\
To define the finite groups that will be used in this work, we must first define the subgroups of the modular group known as congruence subgroups or also known as homogeneous principal congruence groups of level $N$, $\Gamma(N)$. They are defined by imposing a modularity constraint on the entries of the arrays in $\Gamma$, that is,
\begin{align}
    \Gamma(N) =\left\lbrace \left(
        \begin{array}{cc}
            a & b \\ 
            c & d
        \end{array}  
    \right) \in \Gamma :\left(
        \begin{array}{cc}
            a & b \\ 
            c & d
        \end{array}  
    \right)=\left(
        \begin{array}{cc}
            1 & 0 \\ 
            0 & 1
        \end{array}  
    \right)(\text{mod}\ N) \right\rbrace. \label{gcongru}
\end{align}
Similarly, the corresponding inhomogeneous transformation $\gamma(N)$ can be defined, whose associated matrices belong to $\overline{\Gamma}(N)=\Gamma(N)/\{ \mathbf{1}_{2\cross 2},-\mathbf{1}_{2\cross 2}\}$ for $N\leq2$, since the identity matrix $\mathbf{1}_{2\cross 2}$ and $-\mathbf{1}_{2\cross 2}$ are indistinguishable, that is, they lead to the same transformation in $\gamma(N)$. For $N>2$ we have $\overline{\Gamma}(N)=\Gamma(N)$. In the definition (\ref{gcongru}) it is observed that for the particular case $N=1$ the group $\Gamma$ is obtained since all the matrices of $\Gamma$ are equal to the identity matrix $ \bmod\ 1$.\\
Since $\overline{\Gamma}(N)$ is a normal subgroup of $\overline{\Gamma}$, the finite modular group $\Gamma_N$ can be defined as the quotient group
\begin{equation}
    \Gamma_N \equiv \overline{\Gamma}/\overline{\Gamma}(N).
\end{equation} 
It can be shown that there is an isomorphism between some finite modular groups and certain groups of rotations of regular polygons. These isomorphisms are
\begin{align}
    &\Gamma_2 \simeq S_3 \notag \\
    &\Gamma_3 \simeq A_4 \\
    &\Gamma_4 \simeq S_4 \notag \\
    &\Gamma_5 \simeq A_5. \notag 
\end{align}
For $N>5$, isomorphisms to symmetry groups become more complex to find \cite{liu2019neutrino}. The order of different finite modular groups $|\Gamma_N|$ is presented in table \ref{dimgamma}. These orders are determined by the expression
\begin{equation}
    |\Gamma_N|=\frac{1}{2}N^3\prod_{p|N}\left(1-\frac{1}{p^2}\right), \hspace{1cm} N>2,
\end{equation}
with $p$ a prime number.
It is observed that for $\Gamma(N)$, the generator $T$ satisfies
\begin{equation}
    T^N \in \Gamma(N)~,
\end{equation}
since
\begin{equation}
    T^N=\left(
    \begin{array}{cc}
        1 & N \\ 
        0 & 1
    \end{array}  
    \right)=\left(
    \begin{array}{cc}
        1 & 0 \\ 
        0 & 1
    \end{array}  
    \right) (\bmod\ N).
\end{equation}
The arithmetic modularity indicates that
$a=b~(\bmod\ N)$ It is equivalent to $a-b=kN$, with $k$ being some integer. For $T^N$ this condition is satisfied even for the input $(12)$, since $N-0=kN$ if $k=1$.
In this way, a new condition arises on the generators for $\overline{\Gamma}$. Now, they must satisfy
\begin{align}
    S^2 = \mathbf{1} \hspace{1cm} (ST)^3 = \mathbf{1}  \hspace{1cm} T^N = \mathbf{1}. 
\end{align}
There are certain holomorphic functions that depend on the modulus $\tau$, which under the group $\Gamma$ transform as
\begin{equation}
    f(\gamma\tau) = (c\tau+d)^k f(\tau),
\end{equation} 
known as modular forms of weight $k$. Since the identity matrix and its negative are equivalent to the same transformation on $\overline{\Gamma}$, $k$ must be an even number. This is because for $\mathbf{1}$, the modular form transforms as
\begin{align}
    f(\tau)= 1^kf(\tau)
\end{align}
and for $-\mathbf{1}$ transform as
\begin{align}
    f(\tau)= (-1)^kf(\tau).
\end{align}
Since the transformation must be equivalent in both cases, $1^k=(-1)^k$, which is satisfied if $k$ satisfies the property of being even. The above property does not hold in general for $\Gamma(N)$ since $-\mathbf{1}_{2 \cross 2}$ does not belong to $\Gamma(N)$ for $N>2$. Another property of modular forms is that the constant functions in $\mathcal{H}$ are modular forms of zero weight and that zero functions are modular forms of any weight.
For $T^N$, the modular forms transform as
\begin{equation}
    f(T^N\tau) = f(\tau+N) = f(\tau),
\end{equation}
that is, $f(\tau)$ is periodic in $\mathcal{H}$, which allows us to expand in Fourier series \cite{diamond2005first,schoeneberg2012elliptic}
\begin{equation}
    f(\tau)=\sum_{n=0}^{\infty}a_nq^n_N,
\end{equation}
with $q_N=e^{2\pi i \tau/N}$. The set of modular forms forms a vector space denoted by $d_{k}(\Gamma(N))$. The dimension of the vector spaces for some quotient groups $\Gamma_N$, is shown in  Table \ref{dimgamma} \cite{feruglio2019neutrino}. 

\begin{table} 
\tabcolsep 22pt
\centering
    \begin{tabular}{cccc}
    \hline
        $N$ & $d_{k}(\Gamma(N))$ & $|\Gamma_N|$ & $\Gamma_N$\rule[-2ex]{0pt}{5ex}\\
    \hline
        2 & $k/2+1$ & 6 & $S_3$\vspace{0.05cm}\\

        3 & $k+1$ & 12 & $A_4$\vspace{0.05cm}\\

        4 & $2k+1$ & 24 &$S_4$\vspace{0.05cm}\\

        5 & $5k+1$ & 60 & $A_5$\vspace{0.05cm}\\
    \hline
    \end{tabular}
    \caption{Dimension of the vector spaces formed by the quotient groups $\Gamma_N$ as a function of the weight $k$. The third column presents the order of $|\Gamma_N|$ from $N=2$ to $N=5$ \cite{feruglio2019neutrino}.}
    \label{dimgamma}
\end{table}

The Dedekind eta function $\eta(\tau)$ is a special function and is used to construct new modular forms, as will be the case later. This function is defined as
\begin{equation}
    \eta(\tau) = q^{1/24} \prod_{n =1}^\infty (1-q^n),
\end{equation}

with $q=e^{2\pi i \tau}$.

Since the modular forms form a vector space, it can be shown that for a unitary representation of $\Gamma_N$, the modular forms transform as \cite{feruglio2019neutrino,liu2019neutrino}
\begin{equation}
    f(\tau) \rightarrow (c\tau+d)^k\rho(\gamma) f(\tau),
\end{equation}
where $f(\tau)=(f_1(\tau),f_2(\tau),...,f_n(\tau))^T$, with $n=d_{k}(\Gamma(N))$
and $\rho(\gamma)$ are unitary matrices of the representation of the group that satisfy 
\begin{align}
    &\rho(S)^2=\mathbf{1}_{n\cross n} \hspace{0.7cm} \rho(T)^N=\mathbf{1}_{n\cross n} \hspace{0.7cm} \rho(ST)^3=\mathbf{1}_{n\cross n}. 
\end{align}

The importance of modular forms becomes evident within the context of flavor physics, particularly when the different couplings of quarks in the mass matrix are described. {Modular forms have restrictions on their modular weight. However, fields are not modular forms, so their modular weights can be even or negative. The requirement that the Lagrangian remains invariant under the action of a group—specifically $S_3$ in this context—serves to constrain the number of free parameters. Now, the imposition of modular invariance, that is, zero modular weight, imposes additional restrictions on the form that these free parameters can have, particularly in the Yukawa sector. This arises from the necessity for each quark field to be assigned a modular weight such that, together with the modular weight of the forms, gives a total sum equal to zero. Consequently, this requirement limits the number of allowed terms, as not all couplings will be able to meet this criterion. As a result, the quark mass matrix exhibits a structure according to the modular forms, which depend on a single complex parameter $\tau $.

\section{Three Higgs doublets and Modular forms for $S_3$ symmetry} \label{modelo}
In this section, we present the proposed model with three Higgs doubles and with  modular $S_3$ as flavor symmetry.  We will start by introducing the Higgs potential with three electroweak doublets, invariant under $S_3$ and its minimization \cite{Beltran:2009zz,Das:2014fea,Barradas-Guevara:2014yoa,Gomez-Bock:2021uyu}.

\subsection{Higgs potential for three Higgs doublets with \sy symmetry}

For the Higgs sector, the following assignment is considered: the doublets $H_1$ and $H_2$ will be assigned in a doublet of $S_3$ and $H_S$ will be assigned in a symmetric singlet.
The invariance of the Higgs potential under modular $S_3$ can be done if each coupling of the potential is assigned a zero modular weight which, together with the zero weight Higgs doublets, will result in a modular invariant potential. Thus, 
the Higgs potential to be considered must satisfy the symmetry of the group $S_3$. The most general potential under this symmetry is written as follows. 
\begin{align}
    V&=\mu^2_1 \left( H^\dagger_1 H_1 + H^\dagger_2 H_2\right) + \mu^2_0 \left(H^\dagger_S H_S \right) + \frac{a}{2}\left(H^\dagger_S H_S \right)^ 2 \notag\\
    &+ b\left(H^\dagger_S H_S \right)\left( H^\dagger_1 H_1 + H^\dagger_2 H_2\right) \notag \\
    &+\frac{c}{2}\left( H^\dagger_1 H_1 + H^\dagger_2 H_2\right)^ 2 + \frac{d}{2} \left( H^\dagger_1 H_2 - H^\dagger_2 H_1\right)^ 2 \notag \\
    &+ ef_{ijk}\left(\left(H^\dagger_S H_i\right)\left(H^\dagger_j H_k\right) + h.c.\right) \notag \\
    &+f\left\lbrace \left( H^\dagger_S H_1\right)\left( H^\dagger_1 H_S\right) + \left( H^\dagger_S H_2\right)\left( H^\dagger_2 H_S\right) \right\rbrace\\
    &+ \frac{g}{2} \left\lbrace \left( H^\dagger_1 H_1 - H^\dagger_2 H_2\right)^2 + \left( H^\dagger_1 H_2 + H^\dagger_2 H_1\right)^2 \right\rbrace \nonumber\\ 
    &+ \frac{h}{2}\left\lbrace \left( H^\dagger_S H_1\right)\left( H^\dagger_S H_1\right)+ \left( H^\dagger_S H_2\right)\left( H^\dagger_S H_2\right) + \right. \notag
    \\ &\left. \left( H^\dagger_1 H_S\right)\left( H^\dagger_1 H_S\right)  + \left( H^\dagger_2 H_S\right)\left( H^\dagger_2 H_S\right) \right\rbrace; \notag 
\end{align}
where $f_{112} = f_{121} = f_{211} = -f_{222} = 1$ y $a$, $b$, $c$, $d$, $e$, $f$, $g$ and $h$ are self-couplings.

By imposing the condition of the first derivative of the potential equal to zero in the minimization, two of the vacuum expectation values (VEVs) are related by the expression \cite{Das:2014fea,Barradas-Guevara:2014yoa,Gomez-Bock:2021uyu}
\begin{align}
    v_1^2=3v_2^2, \label{relacionp}
\end{align}
where the VEVs are denoted as
\begin{align}
    \bra{0}H_1\ket{0}=\frac{1}{\sqrt{2}}v_1~, ~ 
    \bra{0}H_2\ket{0}=\frac{1}{\sqrt{2}}v_2~, ~ 
    \bra{0}H_s\ket{0}=\frac{1}{\sqrt{2}}v_s~, \notag
\end{align}
and it must be satisfied that $\sqrt{v_1^2+v_2^2+v_s^2}=v=246 \ GeV$. 
It is also possible to write the VEVs in spherical coordinates, as done in \cite{Gomez-Bock:2021uyu},
\begin{eqnarray}
v_1=v\cos\varphi\sin\theta, & v_2= v\sin\varphi\sin\theta, & v_s= v\cos\theta.
\end{eqnarray}
Then, using Eq.~(\ref{relacionp}) they can be rewritten as
\begin{eqnarray}
\tan\varphi=1/\sqrt{3}&\Rightarrow& \sin\varphi=\frac{1}{2},\ \ \  \  \ \ \ \cos\varphi= \frac{\sqrt{3}}{2}, \label{phi}\\
\tan\theta=\frac{2v_2}{v_s}&\Rightarrow& \frac{v}{2}\sin\theta=v_2, \ \ \  \,  \ \ \ v\cos\theta= v_s.\label{angtheta}
\end{eqnarray}

This way the VEVs can be expressed only in terms of one free parameter $\theta$. The relations (\ref{relacionp}) and (\ref{angtheta}) will be useful later to reduce the number of free parameters.

 The minimization of the potential leaves an unbroken $Z_2$ symmetry, subgroup of \sy \cite{Das:2014fea,Gomez-Bock:2021uyu}. In the non-modular \sy version of the model, this leads to zeros in the $V_{CKM}$ matrix. But it also has very interesting features, namely, in the exact SM alignment, where only one of the  scalars is coupled to the gauge bosons and corresponds to the SM Higgs boson, the trilinear and quartic couplings at tree level are exactly the same as in the SM case \cite{Gomez-Bock:2021uyu}. Also, it can  have one or two candidates for dark matter coming from the scalar sector \cite{Gomez-Bock:2021uyu}.  By making \sy modular it is possible to preserve the phenomenologically interesting features of the potential with a realistic  $V_{CKM}$ matrix, as we will show.

\subsection{Construction of modular forms for \sy}  
To proceed, it is necessary to build the modular forms for the $\mathbf{2}$ representation of $S_3$. The procedure for its construction is based on the methods used in \cite{feruglio2019neutrino} by Feruglio and extended to $S_3$ in \cite{kobayashi2019finite} by Kobayashi and is shown in Appendix \ref{apendiceformas}.
A candidate modular form of weight 2 can be constructed from the Dedekind Eta functions, which are defined as
\begin{equation}
    \eta(\tau) = q^{1/24} \prod_{n =1}^\infty (1-q^n),
\end{equation}
where $q = e^{2 \pi i \tau}$. The function $\eta(\tau)$ is a special type of modular form due to its modular weight of $1/2$ and is useful for constructing other modular forms since, under transformations of the generators $S$ and $T$ , a closed algebra is obtained. The transformations under $T$ are
\begin{align}
    &\eta(2\tau) \rightarrow e^{i \pi/6} \eta(2\tau), \nonumber \\
    &\eta(\tau/2) \rightarrow \eta((\tau +1)/2),  \\
    &\eta((\tau + 1)/2) \rightarrow e^{i \pi /12}\eta(\tau/2).  \nonumber 
\end{align}
and under $S$
\begin{align}
    &\eta(2\tau) \rightarrow \sqrt{\frac{-i\tau}{2}} \eta(\tau/2), \nonumber \\
    &\eta(\tau/2) \rightarrow \sqrt{-i3\tau} \eta(2\tau),  \\
    &\eta((\tau + 1)/2) \rightarrow e^{-i\pi/12} \sqrt{-i\tau}\eta((\tau +1)/2). \nonumber  
\end{align}
Since $\Gamma_2\simeq S_3$, it is necessary to use a representation of the generators of $S$ and $T$ in $S_3$ that satisfies $S^2=T^2=(ST)^2=\mathbf{ 1}$. The representation to use is the following
\begin{equation}
    \rho(S) = \frac{1}{2} \left( 
    \begin{array}{cc}
        1 & -\sqrt 3 \\
        -\sqrt 3 & -1
    \end{array}\right), \qquad 
    \rho(T) =  \left( 
    \begin{array}{cc}
        -1 & 0\\
        0 & 1
    \end{array}\right),
\end{equation}
where it can be seen that they are also the generators of $S_3$.
The modular forms of weight 2 for $S_3$ are
\begin{align} 
&Y_1(\tau) = \frac{\sqrt{3}i}{4\pi}\left( \frac{\eta'(\tau/2)}{\eta(\tau/2)}  -\frac{\eta'((\tau +1)/2)}{\eta((\tau+1)/2)}   \right),
    \\
&Y_2(\tau) = \frac{i}{4\pi}\left( \frac{\eta'(\tau/2)}{\eta(\tau/2)}  +\frac{\eta'((\tau +1)/2)}{\eta((\tau+1)/2)} - \frac{8\eta'(2\tau)}{\eta(2\tau)} \right),
\label{formass3}
\end{align}
where the prime indicates the derivative with respect to $\tau$. Figure \ref{Y1Y2} show the real and imaginary parts of the modular forms $Y_1(\tau)$ and $Y_2(\tau)$, respectively. Depending on the assignment made, the modular forms of weight four will be useful and are obtained from the tensor product of the doublet of the modular forms of weight two \cite{kobayashi2019finite}, therefore
\begin{align}
    \left(
    \begin{array}{c}
        Y_1 \\
        Y_2
    \end{array}\right)\otimes\left(
    \begin{array}{c}
        Y_1 \\
        Y_2
    \end{array}
    \right)=Y_s^{(4)}+\left(
    \begin{array}{c}
        Y_1^{(4)} \\
        Y_2^{(4)}
    \end{array}
    \right),
\end{align}
where the antisymmetric singlet vanishes. Furthermore, it has been defined
\begin{align}
    &Y_s^{(4)}=Y_1^2+Y_2^2 \hspace{1cm} Y_1^{(4)}=2Y_1Y_2 \hspace{1cm} Y_2^{(4)}=Y_1^2-Y_2^2. 
\end{align}

\begin{figure*}[ht]
	\centering 
	\includegraphics[scale=0.35, angle=0]{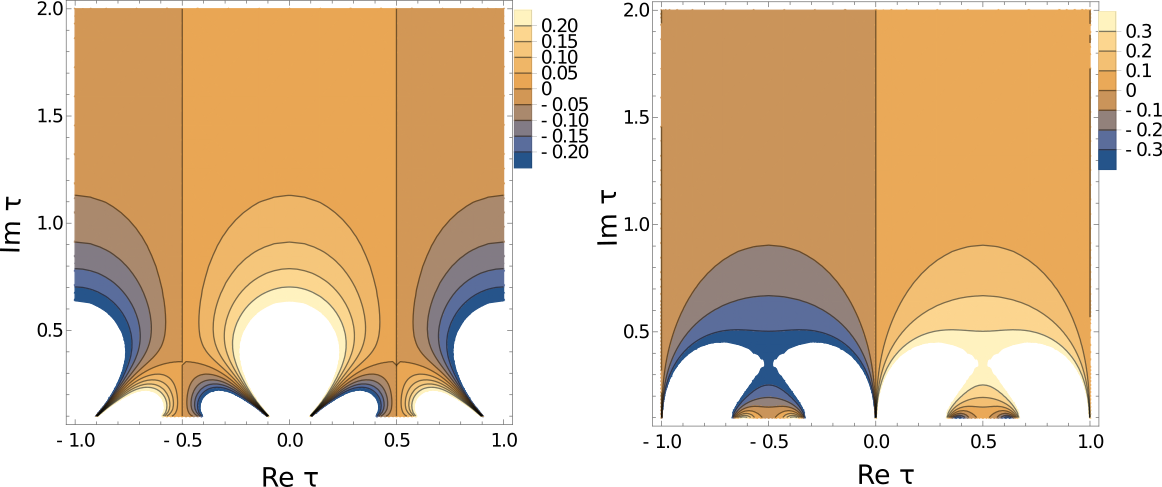}	
    \includegraphics[scale=0.35]{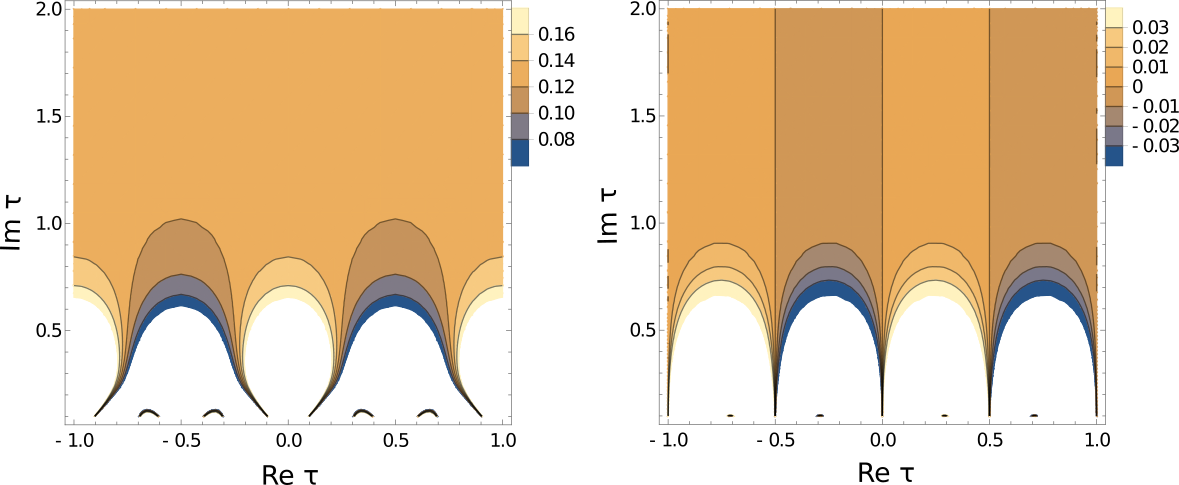}
    \caption{Real (top-left) and imaginary (top-right) part of the modular form $Y_1(\tau)$. Real (bottom-left) and imaginary (bottom-right) part of the modular form $Y_2(\tau)$. The orange color indicates values that tend towards positive and the blue color indicates values that tend towards negative. Blanks are part of the cut when calculating very small or very large amounts.}
    \label{Y1Y2}
\end{figure*}

\subsection{Assignments under \sy and mass matrix}
So far we have introduced the elements to build a model of three Higgs doublets under modular $S_3$ symmetry, that is, $SU(3)_C \cross SU_L(2) \cross U_y(1) \cross \Gamma_2$. It is important to mention that it is going to be assumed that the transformation under the modular group must also be imposed for the quark fields in order to assign them a modular weight, that is,
\begin{equation}
    \phi\rightarrow (c\tau+d)^{k_\phi}\phi,
\end{equation}
where $k_{\phi}$ is the modular weight. It should be clarified that this assumption could be motivated if the modular symmetry of quarks is considered as a residual symmetry of a more fundamental one in the low energy limit. Since the fields are not modular forms, there is the freedom to choose the modular weight in their transformation to obtain an appropriate mass matrix. 

\begin{figure*}[ht]
    \centering
    \includegraphics[scale=0.35]{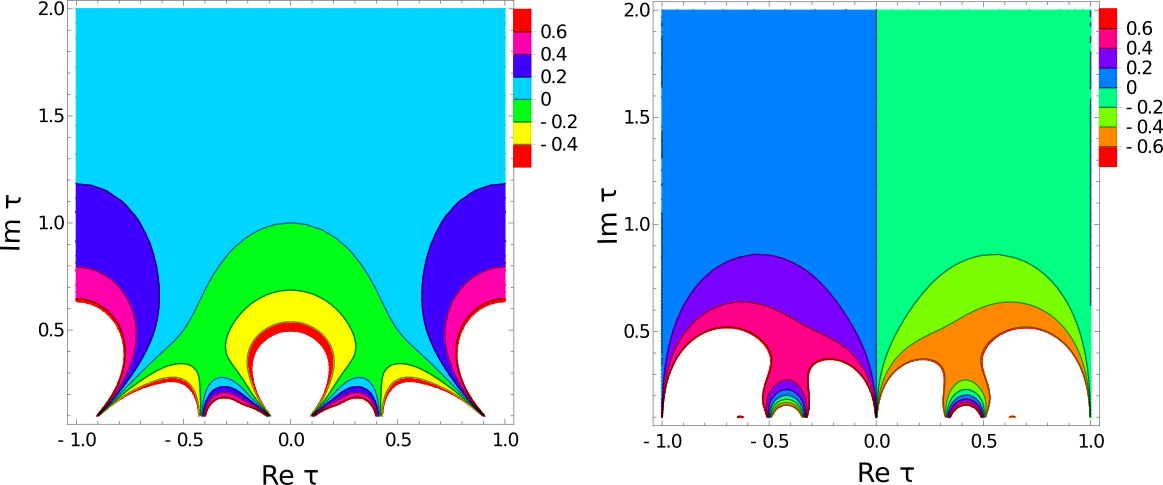}
    \caption{Real (left) and imaginary (right) part of the given expression in $M_{13}$ y $M_{31}$, that is, $Y_2^{(2)}(\tau)-\sqrt{3}Y_1^{(2)}(\tau)$. It is observed that $Y_2^{(2)}(\tau)-\sqrt{3}Y_1^{(2)}(\tau)=0$, for both its real and imaginary parts, at the point $\tau=i$, which guarantees that $M_{13}=M_{31}=0$.}
    \label{Y3}
\end{figure*}

In addition to the quark and Higgs fields, there will be modular forms that, in this case, will be of weight two and four, adding more restrictions on the form of the Yukawa sector of the Lagrangian. The assignment of the quark and Higgs fields for $S_3$ is done in such a way that the sum of the modular weights is zero, which implies that the Lagrangian is modular invariant. There are different options for the assignment, despite this, there are limited assignments that avoid an excess of free parameters or that lead to matrices that are  useful for model building. 

The proposed assignment is as follows: the left electroweak doublets of the first and second families, $Q_1$ and $Q_2$, will be a doublet in $S_3$, likewise, the up- and down-type right singlets, $u_{1R }$, $u_{2R}$ and $d_{1R}$, $d_{2R}$. The left electroweak doublet of the third family, $Q_3$, and the right singlets of the third family, $u_{3R}$ and $d_{3R}$ will be singlets of $S_3$. For the Higgs doublets, two of them, $H_1$ and $H_2$, will form a $S_3$ doublet, while the third, $H_s$, will be a symmetric singlet. Regarding modular weights, all doublets of $S_3$ will have weight $-2$ and singlets and Higgses will have weight 0. These assignments are summarized in  Table \ref{asig}. This assignment allows us to write the Lagrangian in the Yukawa sector as in Eq.(\ref{tensor}), where modular forms of weight four have been used. To compress the notation, the doublets of $S_3$ can be written as
\begin{eqnarray}
    Q=\left(
    \begin{array}{c}
        \overline{Q}_1  \\
        \overline{Q}_2
    \end{array}\right); \ \
    u=\left(
    \begin{array}{c}
        u_{1R}  \\
        u_{2R}
    \end{array}\right); \ \
    H=\left(
    \begin{array}{c}
        H_1  \\
        H_2
    \end{array}\right); \notag \\
    \\
    Y^{(4)}=
    \left(
    \begin{array}{c}
        Y^{(4)}_1 \\
        Y^{(4)}_2
    \end{array}\right); \ \ 
    Y^{(2)}=
    \left(
    \begin{array}{c}
        Y^{(2)}_1 \\
        Y^{(2)}_2
    \end{array}\right)~. \notag 
\end{eqnarray}
Thus, the Lagrangian in the Yukawa sector is
\begin{align}
   \label{tensor} \mathcal{L}_y^{(u)} &=
    C_1\overline{Q}\otimes u\otimes \tilde{H} \otimes Y^{(4)}
    +
    C_2\overline{Q}
    \otimes
    u
    \otimes
    \tilde{H}
    \otimes
    Y_s^{(4)} \notag\\&+
    C_3\overline{Q}
    \otimes
    u
    \otimes
    \tilde{H}_s
    \otimes
    Y^{(4)}  
    +
    C_4\overline{Q}
    \otimes
    u
    \otimes
    \tilde{H}_s
    \otimes
    Y^{(4)}_s \notag 
    \\ &+C_5\overline{Q}
    \otimes
    u_{3R}
    \otimes
    \tilde{H}
    \otimes
    Y^{(2)}
    +
    C_6\overline{Q}
    \otimes
    u_{3R}
    \otimes
    \tilde{H}_s
    \otimes
    Y^{(2)} 
    \\&+C_7
    \overline{Q}_3
    \otimes
    u
    \otimes
    \tilde{H}
    \otimes
    Y^{(2)}
    +
    C_8
    \overline{Q}_3
    \otimes
    u
    \otimes
    \tilde{H}_s
    \otimes
    Y^{(2)}\notag 
    \\&+C_9
    \overline{Q}_3 
    \otimes
    u_{3R}
    \otimes
    \tilde{H}_s + \text{h.c.}\notag 
\end{align}
\begin{table*}[ht]
    \tabcolsep 10pt
    \vspace{1mm}
    \begin{tabular}{ccccccccc}
        \hline 
        & $(Q_1,Q_2)$ & $(q_1,q_2)$ & $Q_3$ & $q_3$ & $(H_1,H_2)$ & $H_s$ & $(Y_1^{(2,4)}(\tau),Y_2^{(2,4)}(\tau))$ & $Y_s^{(4)}(\tau)$\rule[-2ex]{0pt}{5ex}\\ 
        \hline 
        $SU(2)$ & $2$ & $1$ & $2$ & $1$ & $2$ & $2$ & $1$ & $1$ \vspace{0.05cm}\\ 
        $S_3$ & $2$ & $2$ & $1$ & $1$ & $2$ & $1$ & $2$ &  $1$ \vspace{0.05cm}\\ 
        $k$ & $-2$ & $-2$ & $0$ & $0$ & $0$ & $0$ & $(2,4)$ & $4$ \vspace{0.05cm}\\
        \hline
    \end{tabular} 
    \caption{charges, assignments, and modular weights of $SU(2)$ and $S_3$. The superscript $(2,4)$ on the modular forms indicates that they are of modular weight 2 or 4. The subscript $s$ indicates the symmetric singlet of the modular form of weight 4.}
    \label{asig}
\end{table*}
The expanded Lagrangian $\mathcal{L}_y^{(u)}$ is shown in Appendix \ref{complagrangian}. 

After spontaneous symmetry breaking, the matrix elements $M_{ij}^{(u)}$ are determined by
\begin{align}
&M_{11}^{(u)}=(\alpha+\gamma)v_1Y_1^{(4)}+(\alpha-\gamma)v_2Y_2^{(4)}+C_2v_2Y_s^{(4)}+C_3v_sY_2^{(4)}+C_4v_sY_s^{(4)} \notag \\ 
&M_{12}^{(u)}=(\beta+\gamma)v_2Y_1^{(4)}+(\gamma-\beta)v_1Y_2^{(4)}+C_2v_1Y_s^{(4)}+C_3v_sY_1^{(4)} \notag \\ 
&M_{13}^{(u)}=C_5(v_2Y_1^{(2)}+v_1Y_2^{(2)})+C_6v_sY_1^{(2)}\notag \\
&M_{21}^{(u)}=(\beta+\gamma)v_1Y_2^{(4)}+(\gamma-\beta)v_2Y_1^{(4)}+C_2v_1Y_s^{(4)}+C_3v_sY_1^{(4)} \notag \\
&M_{22}^{(u)}=(\alpha+\gamma)v_2Y_2^{(4)}+(\alpha-\gamma)v_1Y_1^{(4)}-C_2v_2Y_s^{(4)}-C_3v_sY_2^{(4)}+C_4v_sY_s^{(4)} \notag \\ 
&M_{23}^{(u)}=C_5(v_1Y_1^{(2)}-v_2Y_2^{(2)})+C_6v_sY_2^{(2)} \\ \notag
&M_{31}^{(u)}=C_7(v_2Y_1^{(2)}+v_1Y_2^{(2)})+C_8v_sY_1^{(2)}\notag \\
&M_{32}^{(u)}=C_7(v_1Y_1^{(2)}-v_2Y_2^{(2)})+C_8v_sY_2^{(2)}\notag \\ 
&M_{33}^{(u)}=C_9v_s, \notag
\end{align} 
with $v_1$, $v_2$ y $v_s$ the VEVs of the Higgs fields. The parameters $\alpha$, $\beta$, $\gamma$ arise as a result of the invariant product in the $C_1$ term in \ref{tensor}. In this model, the free parameters are $\alpha$, $\beta$, $\gamma$, $v_2$, $v_s$ $C_2$, $C_3$, $C_4$, $C_5$, $C_6$, $C_7$, $C_8$, $C_9$ y $\tau$, even so, the matrix reflects some symmetries between its components, which will allow reducing or eliminating some of them.
Similarly, for the matrix $M_{ij}^{(d)}$. $C_9$ can be considered a real parameter, thus, there are a total of 47 independent real parameters.

\section{Calculation of the texture-zero mass matrix}
\label{calculationsmass}
It is desirable to obtain  texture zeros in the quark mass matrix since by doing so a suitable $V_{CKM}$ can be obtained \cite{FRITZSCH20001}. The texture we aim to obtain has the following form, which has been shown to lead to a phenomenologically viable $V_{CKM}$ \cite{Xing:2020ijf}
\begin{align}
    \left(
    \begin{array}{ccc}
        0     & a & 0 \\
        a^* & b & c \\
        0 & c^* & d
    \end{array} \label{matriztex}
    \right) ~,
\end{align}
which is hierarchical, meaning $d \gg c \gtrsim b \gg a$.
To satisfy this form in the mass matrix, we must impose the conditions
\begin{align}
    M_{11}=0 \hspace{0.7cm}
    M_{12}=M_{21}^* \hspace{0.7cm}
    M_{32}=M_{23}^* \hspace{0.7cm}
    M_{13}=M_{31}=0 ~,  
    \label{condim}
\end{align}
which can be achieved through a rotation and a shift (see for instance \cite{canales2013quark}).

To guarantee the existence of a transformation that allows one to obtain the conditions in \ref{condim}, a rotation by an angle $\theta$ is performed on the component $M^{(u)}_{33}$,
\begin{align}
    M^{(u)}_R = R^T M^{(u)} R,
\end{align}
where $M^{(u)}_R$ is the rotated matrix and $R$ is the rotation matrix.

The most problematic conditions are $M_{13} = M_{31} = 0$. By setting $\tau = i$, we obtain the relation
\begin{equation}
Y_2^{(2)}(i) - \sqrt{3}\, Y_1^{(2)}(i) = 0,
\end{equation}
which is a useful identity (see Eq.~\ref{relacionformas} in Appendix~\ref{apendiceformas}). Using also the relation $v_1^2 = 3 v_2^2$, which arises from the minimization of the scalar potential, these conditions can be written as
\begin{align}
    (4 C_5 v_2 + C_6 v_s)\cos\vartheta + \sqrt{3} C_6 v_s \sin\vartheta &= 0, \\
    (4 C_7 v_2 + C_8 v_s)\cos\vartheta + \sqrt{3} C_8 v_s \sin\vartheta &= 0.
\end{align}

These expressions can be rewritten as
\begin{align}
    \tan\vartheta &= -\frac{4 C_5 v_2 + C_6 v_s}{\sqrt{3} C_6 v_s} 
    = -\frac{4 C_7 v_2 + C_8 v_s}{\sqrt{3} C_8 v_s}.
\end{align}

Since an angle $\vartheta$ exists for any combination of parameters, as long as $C_6$ and $C_8$ are different from zero, it is therefore possible to find different configurations that satisfy the conditions in \ref{condim}, which then lead to relations among some of the parameters, namely $C_5C_8 = C_6C_7$. Two setups will be presented, in which these relationships are satisfied, and for simplicity we just set the values for the rest of the parameters.

\subsection{Setup 1}
\label{caso1}
In the first setup, the texture matrix \ref{matriztex} is obtained by setting 
\begin{align*}
    \begin{array}{ll}
        Re(\beta) =0 & \hspace{1cm}C_3=0 \\
        C_4=0 & \hspace{1cm}C_5=C_7^* \\
        C_6=-4(v_2/v_s)C_5 & \hspace{1cm}C_8=-4(v_2/v_s)C_7 \\
        \alpha= -C_2 \in \mathds{R} & \hspace{1cm}\tau=i \\
        \gamma=0 & \hspace{1cm}C_9,v_{1,2},v_s \in \mathds{R}
    \end{array}
\end{align*}
If we put
\begin{align}
    &y_2=\sqrt{3}y_1^2 \hspace{1cm} y_1^{(4)}=2\sqrt{3}y_1^2\notag \\
    &y_2^{(4)}=-2y_1^2 \hspace{1cm} y_s^{(4)}=4y_1^2,
\end{align}
with $y_k=Y_k(i)$, the mass matrix takes the form
\begin{align}
    \hat{M}^{(u)}=\left(
    \begin{array}{ccc}
        0 & C_2'+C_\beta' & 0 \\
        C_2'-C_\beta' &-\frac{2}{\sqrt{3}}C_2' & C_5' \\
        0 & C_5^{'*} & C_9'
    \end{array}
    \right),
\end{align}
where
\begin{align}
    &C_2' = 4\sqrt{3}v_2y_1^2C_2, \hspace{1cm} C_\beta' = 4\sqrt{3}v_2y_1^2\beta, \notag \\
    &C_5'=-4\sqrt{3}v_2y_1C_5, \hspace{1cm} C_9'=C_9vs ~. 
\end{align}
In Refs.~\cite{canales2013quark,barranco2010universal} a method is presented to diagonalize this type of texture, as well as analytical expressions for matrix elements $V_{CKM}$ in terms of mass ratios. Following this procedure, the matrix $M^{(u)}$ can be rewritten in polar form in the following way
\begin{align}
    P_f = \textit{diag}(1, e^{i\phi_{1u}}, e^{i(\phi_{1u}-\phi_{2u})})
\end{align}
where $\phi_{1u}$ is the phase of $C_2+C_\beta$ and $\phi_{2u}$ is the phase of $C_5$. Therefore, a new matrix, $\bar{M}^{(u)}$, is defined such that
\begin{align}
    M^{(u)}=P_f^{\dagger}\bar{M}^{(u)}P_f,
\end{align}
Thus,
\begin{align}
    \bar{M}^{(u)}=
    \left(
    \begin{array}{ccc}
        0 & |C| & 0 \\
        |C| & -\frac{2}{\sqrt{3}}|C|\cos(\phi_{1u}) & |C_5'| \\
        0 & |C_5'| & C_9'
    \end{array}
    \right),
\end{align}
with $C=C_2'+C_\beta'$ y $\cos\phi_{1u}=C_2'/C$. It should be noted that there are a total of five parameters that describe the matrix are: $|C|$, $|C_5'|$, $|C_9'|$, $\phi_{1u}$ y $\phi_{2u}$.
\begin{table*}
    \tabcolsep 25pt
    \begin{center}
        \begin{tabular}{ccc}
            \hline 
            Quarks & Masses $[GeV]$ (PDG) (2022) & Masses in $\overline{MS}$ ($M_Z$) $[GeV]$ \rule[-2ex]{0pt}{5ex}\\ 
            \hline \\
            $m_u$ & $2.16^{+0.49}_{-0.26} \cross 10^{-3}$   & $0.0012^{+0.0003}_{-0.0001}$\vspace{0.05cm} \\
            $m_d$ & $4.67^{+0.48}_{-0.17} \cross 10^{-3}$    & $0.0027^{+0.0003}_{-0.0001}$ \vspace{0.05cm}\\
            $m_s$ & $93.4^{+8.6}_{-3.4}\cross 10^{-3}$  & $0.0545^{+0.0050}_{-0.0019}$  \vspace{0.05cm}\\
            $m_c$ & $1.27$ $\pm$ $0.02$  & $0.647 \pm 0.010$  \vspace{0.05cm}\\
            $m_b$ & $4.18^{+0.03}_{-0.02}$  & $2.86^{+0.02}_{-0.001}$  \vspace{0.05cm}\\
            $m_t$ & $172.69$ $\pm$ $0.30$  & $170.63 \pm 0.07$ \vspace{0.05cm}\\
            \hline 
        \end{tabular} 
        \caption{Information on the quark masses of the PDG (2022) \cite{Workman:2022ynf} and the masses in the $\overline{MS}$ scheme at the scale of $M_Z$. Masses were calculated at four loops for both $\alpha_s$ and masses. The package used was RunDec for Mathematica.}
        \label{tablamasas}
    \end{center}
\end{table*}
Using the mathematical properties of matrix invariants, specifically the trace, determinant, and the trace of the square of the matrix, it is possible to reparameterize the mass matrices using mass ratios. We define the normalized matrix as $\bar{M}^{(u)}_n=\bar{M}^{(u)}/m_3$, alongside the redefinitions $|K|=|C|/m_3$, $|K_5|=|C_5'|/m_3$, and $K_9=C_9'/m_3$.
 The eigenvalues of these mass matrices can  be expressed in terms of the ratios of the first two generations and the third, which is the most massive. 
This is expressed as $\widetilde{\sigma}_i=m_i/m_3$ for both up-type and down-type quarks, as detailed in Table \ref{tablamasas} \footnote{The values presented in Table \ref{tablamasas} and Table \ref{tablamedidas}  have not changed significantly with respect to the PDG 2025 values} . Consequently, the mass matrix in its diagonal form is represented as $M_D=\text{diag}(\widetilde{\sigma}_1,-\widetilde{\sigma}_2,1)$. The minus sign in $\widetilde{\sigma}_2$ is required in order for this parameter remains a positive real number. This is obtained through the chiral transformations $\psi'_{L}=e^{-i\gamma5\frac{\pi}{2}\psi_L}$ and $\psi'_{R}=e^{i\gamma5\frac{\pi}{2}\psi_R}$. From the invariants in $\bar{M}^{(u)}_n$ and $M_D$, we obtain the following relationships:
\begin{align}
    |K|&=\sqrt{\frac{\widetilde{\sigma}_1\widetilde{\sigma}_2}{K_9}}\notag \\ 
    \cos\phi_{1u}&=\frac{\sqrt{3}}{2}(K_9-\widetilde{\sigma}_1+\widetilde{\sigma}_2-1)\sqrt{\frac{K_9}{\widetilde{\sigma}_1\widetilde{\sigma}_2}} 
    \\
    |K_5|&=\sqrt{\frac{(1-K_9)(K_9-\widetilde{\sigma}_1)(K_9+\widetilde{\sigma}_2)}{K_9}}, \notag
\end{align}
Similarly,  the  down type matrix contains only two free parameters. Therefore, the total number of free parameters in the mass matrices are four: $K_{9u}$, $K_{9d}$, $\phi_{2u}$ and $\phi_{2d}$.  Despite the presence of additional parameters in the mass matrix, they are constrained greatly by the previous relations. But this particular parameterization does not meet the conditions for a good texture for the mass matrix, since the 22 term will always be smaller than the 12, and will not have the hierarchy among parameters of Eq.(\ref{matriztex}),  preventing a proper fit to the expected CKM matrix, as discussed in Section \ref{vckmmatrix}.

\subsection{Setup 2}
A different definition of the parameters allows building a model with more parameters than the previous one, in this way, there is more freedom to adjust these parameters to obtain an appropriate $V_{CKM}$ matrix. Again, the conditions given in \ref{condim} are satisfied if
\begin{align*}
    \begin{array}{lll}
    Re(\beta) =0 & \hspace{0.2cm} C_3=0 & \hspace{0.2cm}\gamma = -(1/2)(v_s/v_2)C_4 \in \mathds{R} \\
    \alpha= -C_2 \in \mathds{R} & \hspace{0.2cm}C_6=-4(v_2/v_s)C_5 & \hspace{0.2cm}C_8=-4(v_2/v_s)C_7 \\
    C_5=C_7^* & \hspace{0.2cm}\tau=i & \hspace{0.2cm}C_9,v_{1,2},v_s \in \mathds{R}
    \end{array}
\end{align*}

In the case of the condition $M_{13}=M_{31}=0$, the equality is satisfied by using the conditions on $C_6$ and $C_8$, the relation $v^2_1=3v_2^2$ that arises from the minimization of the potential, and $Y_2^{(2)}(\tau)-\sqrt{3}Y_1^{(2)}(\tau)=0$  at $\tau=i$, thus
\begin{align}
    &y_2=\sqrt{3}y_1, \hspace{1cm}y_1^{(4)}=2\sqrt{3}y_1^2,\\ \notag
    &y_2^{(4)}=-2y_1^2, \hspace{1cm} y_s^{(4)}=4y_1^2, 
\end{align}
with $y_k=Y_k(i)$. The mass matrix takes the form
\begin{align}
    M^{(u)}=
    \left(
    \begin{array}{ccc}
        0 & C' & 0 \\
        C^{'*} & C'_4  & C'_5 \\
        0 & C_5^{'*}  & C'_9
    \end{array}
    \right),
\end{align}
where
\begin{align}
    &C' = 4\sqrt{3}v_2y_1^2(C_2+\beta) = 2\sqrt{3}v \sin\theta y_1^2(C_2+\beta) , \notag \\
    &C'_4=8 y_1^2 (C_4 v_s-C_2 v_2) = 4 y_1^2 v (2C_4 \cos\theta - C_2\sin\theta),  \\
    &C'_5=-4\sqrt{3}v_2y_1C_5 = -2\sqrt{3} y_1C_5 v\sin\theta~,\notag \\
    &C_9'=C_9v_s = C_9v\cos\theta \notag
    \label{eq:Csup}
\end{align}
where the geometrical parameterization of the VEVs Eq.(\ref{angtheta}) was used. Following  a procedure similar to the previous setup, the following mass matrix is obtained
\begin{align}
    \bar{M}^{(u)}=
    \left(
    \begin{array}{ccc}
    0 & |C| & 0 \\
    |C| & C_4' & |C_5'| \\
    0 & |C_5'| & C_9'
    \end{array}
    \right),
\end{align}
with the following relations between its components by means of the three matrix invariants already mentioned,
\begin{align}
    &|K|=\sqrt{\frac{\widetilde{\sigma}_1\widetilde{\sigma}_2}{K_9}}\notag \\
    &K_4=(\widetilde{\sigma}_1-\widetilde{\sigma}_2+1-K_9) 
    \\
    &|K_5|=\sqrt{\frac{(1-K_9)(K_9-\widetilde{\sigma}_1)(K_9+\widetilde{\sigma}_2)}{K_9}}, \notag
\end{align}
with $|K|=|C|/m_3$, $|K_5|=|C_5'|/m_3$, $K_4=C'_4/m_3$ and $K_9=C_9'/m_3$. In this way, all elements of the array can be described with a single parameter, $K_9$. It should be noted that this procedure is analogous for the case $M^{(d)}$, therefore, the total free parameters remaining for the mass matrices are: $K_{9u}$, $K_{9d}$ , $\phi_{1u}$,$\phi_{2u}$, $\phi_{1d}$ and $\phi_{2d}$.
It is also possible to set one of the phases of each sector to zero without loss of generality, which means each mass matrix is determined by one real parameter  and one phase.

\subsection{The $V_{CKM}$ matrix}
\label{vckmmatrix}
Expressions for the matrix elements of $V_{CKM}$ are presented in \cite{canales2013quark,barranco2010universal}. The elements are calculated with the matrices, $A_L$ and $A_R$, that diagonalize the mass matrix. These matrices must be unitary, therefore, we can build a system of equations for each of the matrices. With the definition of the quantities
\begin{align}
    \label{Ds}
    &\delta_{u,d}=1-K_{9u,d} \notag \\
    &\xi_{1}^{u,d} = 1 - \widetilde{\sigma}_{u,d} - \delta_{u,d} , \quad \notag \\
    &\xi_{2}^{u,d} = 1 + 
    \widetilde{\sigma}_{c,s} - \delta_{u,d},\nonumber\\
    &{\cal D}_{ 1(u,d) } = ( 1 - \delta_{u,d} )( \widetilde{\sigma}_{u,d} + \widetilde{\sigma}_{c,s} )( 1 - 
    \widetilde{\sigma}_{u,d} ), \\
    & {\cal D}_{ 2(u,d) } = ( 1 - \delta_{u,d} )( \widetilde{\sigma}_{u,d} + \widetilde{\sigma}_{c,s} )( 1 + 
    \widetilde{\sigma}_{c,s} ),  \nonumber\\
    &{\cal D}_{3(u,d)} = ( 1 - \delta_{u,d} )( 1 - \widetilde{\sigma}_{u,d} )( 1 + \widetilde{\sigma}_{c,s} ). \notag
\end{align}	  
The elements of the mixing matrix are expressed as \footnote{These expressions have corrections in typos with respect to those presented in \cite{canales2013quark,barranco2010universal}}
{\scriptsize
\begin{align}
\label{elem:ckm_S3SM}
    &V_{ ud }^{ ^{th} } = 
    \sqrt{ \frac{ \widetilde{\sigma}_{c} \widetilde{\sigma}_{s} \xi_{1}^u  \xi_{1}^d }{ 
    {\cal D}_{ 1u } {\cal D}_{ 1d } } } 
    + \sqrt{ \frac{ \widetilde{\sigma}_{u} \widetilde{\sigma}_{d} }{ 
    {\cal D}_{ 1u } {\cal D}_{ 1d } } } \left( \sqrt{ \left( 1 - \delta_{ u } \right) 
    \left( 1 - \delta_{d} \right) \xi_{ 1 }^u \xi_{ 1 }^d } + \sqrt{ \delta_{u} \delta_{d} \xi_{ 2 }^u 
    \xi_{ 2 }^d }e^{ i \phi_2 } \right) e^{ i \phi_1 }, \notag \\
    &V_{us}^{ ^{th} } = 
    -\sqrt{ \frac{ \widetilde{\sigma}_{c} \widetilde{\sigma}_{d} \xi_{ 1 }^u \xi_{ 2 }^d }{ 
    {\cal D}_{ 1u } {\cal D}_{ 2d } } } + \sqrt{ \frac{ \widetilde{\sigma}_{u} \widetilde{\sigma}_{s} }{ 
    {\cal D}_{ 1u } {\cal D}_{ 2d } } } \left( \sqrt{ \left( 1 - \delta_{u} \right) \left( 1 - 
    \delta_{d} \right) \xi_{ 1 }^u \xi_{ 2 }^d} + \sqrt{ \delta_{u} \delta_{d} \xi_{ 2 }^u \xi_{ 1 }^d }e^{ i \phi_2 } 
    \right) e^{ i \phi_1 }, \notag\\
    &V_{ub}^{ ^{th} } = 
    \sqrt{ \frac{ \widetilde{\sigma}_{c} \widetilde{\sigma}_{d} \widetilde{\sigma}_{s} \delta_{d} \xi_{ 1 }^u }{ 
    {\cal D}_{ 1u } {\cal D}_{ 3d } } } + \sqrt{ \frac{ \widetilde{\sigma}_{u} }{ 
    {\cal D}_{ 1u } {\cal D}_{ 3d } } } \left( \sqrt{ \left( 1 - \delta_{u} \right) \left( 1 - 
    \delta_{d} \right) \delta_{d} \xi_{ 1 }^u } - \sqrt{ \delta_{u} \xi_{ 2 }^u \xi_{ 1 }^d \xi_{ 2 }^d }e^{ i \phi_2 } 
    \right) e^{ i \phi_1 }, \notag\\
    &V_{cd}^{ ^{th} } = 
    -\sqrt{ \frac{ \widetilde{\sigma}_{u} \widetilde{\sigma}_{s} \xi_{ 2 }^u \xi_{ 1}^d }{ 
    {\cal D}_{ 2u } {\cal D}_{ 1d } } } + \sqrt{ \frac{ \widetilde{\sigma}_{c} \widetilde{\sigma}_{d} }{
    {\cal D}_{ 2u } {\cal D}_{ 1d } } } \left( \sqrt{ \left( 1 - \delta_{u} \right) \left( 1 - 
    \delta_{d} \right) \xi_{ 2 }^u \xi_{ 1 }^d } + \sqrt{ \delta_{u} \delta_{d} \xi_{ 1 }^u \xi_{ 2 }^d }e^{ i \phi_2 }  
    \right) e^{ i \phi_1 }, \notag \\
    &V_{cs}^{ ^{th} } = 
    \sqrt{ \frac{ \widetilde{\sigma}_{u} \widetilde{\sigma}_{d} \xi_{ 2 }^u \xi_{ 2}^d }{ 
    {\cal D}_{ 2u } {\cal D}_{ 2d } } } + \sqrt{ \frac{ \widetilde{\sigma}_{c} \widetilde{\sigma}_{s} }{
    {\cal D}_{ 2u } {\cal D}_{ 2d } } } \left( \sqrt{ \left( 1 - \delta_{u} \right) \left( 1 - 
    \delta_{d} \right) \xi_{ 2 }^u \xi_{ 2 }^d } + \sqrt{ \delta_{u} \delta_{d} \xi_{ 1 }^u \xi_{ 1 }^d }e^{ i \phi_2 }  
    \right) e^{ i \phi_1 }, \\
    &V_{cb}^{ ^{th} } = 
    -\sqrt{ \frac{ \widetilde{\sigma}_{u} \widetilde{\sigma}_{d} \widetilde{\sigma}_{s} \delta_{d} \xi_{ 2 }^u 
    }{ {\cal D}_{ 2u } {\cal D}_{ 3d } } } + \sqrt{ \frac{ \widetilde{\sigma}_{c} }{ 
    {\cal D}_{ 2u } {\cal D}_{ 3d } } }  \left( \sqrt{ \left( 1 - \delta_{u} \right) \left( 1 
    -\delta_{d} \right) \delta_{d} \xi_{ 2 }^u } - \sqrt{ \delta_{u} \xi_{ 1 }^u \xi_{ 1 }^d \xi_{ 2 }^d 
    }e^{ i \phi_2 } \right) e^{ i \phi_1 } , \notag \\
    &V_{td}^{ ^{th} } = 
    \sqrt{ \frac{ \widetilde{\sigma}_{u} \widetilde{\sigma}_{c} \widetilde{\sigma}_{s} \delta_{u} \xi_{ 1 }^d }{ 
    {\cal D}_{ 3u } {\cal D}_{ 1d } } } + \sqrt{ \frac{ \widetilde{\sigma}_{d} }{ 
    {\cal D}_{ 3u } {\cal D}_{ 1d } } } \left( \sqrt{ \delta_{u} \left( 1 - \delta_{u} \right) 
    \left( 1 - \delta_{d} \right) \xi_{ 1 }^d } - \sqrt{ \delta_{d} \xi_{ 1 }^u \xi_{ 2 }^u \xi_{ 2 }^d }e^{ i \phi_2 } 
    \right) e^{ i \phi_1 },\notag \\
    &V_{ts}^{ ^{th} } = 
    -\sqrt{ \frac{ \widetilde{\sigma}_{u} \widetilde{\sigma}_{c} \widetilde{\sigma}_{d} \delta_{u} \xi_{ 2}^d }{ 
    {\cal D}_{ 3u } {\cal D}_{ 2d } } } + \sqrt{ \frac{ \widetilde{\sigma}_{s} }{ 
    {\cal D}_{ 3u } {\cal D}_{ 2d } } } \left( \sqrt{ \delta_{u} \left( 1 - \delta_{u} \right) 
    \left( 1 - \delta_{d} \right) \xi_{ 2 }^d } - \sqrt{ \delta_{d} \xi_{ 1 }^u \xi_{ 2 }^u \xi_{ 1 }^d }e^{ i \phi_2 } 
    \right) e^{ i \phi_1 },\notag\\
    &V_{tb}^{ ^{th} } = 
    \sqrt{ \frac{ \widetilde{\sigma}_{u} \widetilde{\sigma}_{c} \widetilde{\sigma}_{d} \widetilde{\sigma}_{s} 
    \delta_{u} \delta_{d} }{ {\cal D}_{ 3u } {\cal D}_{ 3d } } } + \left( \sqrt{ 
    \frac{ \xi_{ 1 }^u \xi_{ 2 }^u \xi_{ 1 }^d \xi_{ 2 }^d }{ {\cal D}_{ 3u } {\cal D}_{ 3d } } } 
    +\sqrt{ \frac{ \delta_{u} \delta_{d} \left( 1 - \delta_{u} \right) \left( 1 - \delta_{d} 
    \right) }{ {\cal D}_{ 3u } D_{ 3d } } }e^{ i \phi_2 } \right) e^{ i \phi_1 } ~. \notag
 \end{align}
 }
Where $\phi_i=\phi_{iu}-\phi_{id}$.
 
 For the model analysis, the following $\chi^2$ function will be used 
 \begin{align}
 \chi^2=\sum_{i=u,c,t}\sum_{j=d,s,b}\frac{\left(|V_{ij}^{\text{th}}|-|V_{ij}|\right)^2}{\sigma_{V_{ij}}^2}
 \end{align}
where  $\sigma$ is the corresponding uncertainty. This analysis allows us to obtain the values of the parameters that minimize $\chi^2$. Setup 1 shown in \ref{caso1} will not be analyzed due to strong phase constraints, which leads to the parameters not having the required hierarchy, as already mentioned. Since $\phi_i=\phi_{iu}-\phi_{id}$, the total number of parameters is 3, which made it impossible to properly fit the $V_{CKM}$ matrix. Thus, only Setup 2 will be analyzed. \\
\begin{table}[ht]
\tabcolsep 10pt
\begin{center}
\begin{tabular}{cc}
\hline 
Matrix elements & Numerical value (PDG) (2022)\rule[-2ex]{0pt}{5ex} \\
\hline \\
 $|V_{ud}|$ & $0.97435\pm 0.00016$ \vspace{0.05cm}\\
 $|V_{us}|$ & $0.22500\pm 0.00067$ \vspace{0.05cm}\\
 $|V_{ub}|$ & $0.00369\pm 0.00011$ \vspace{0.05cm}\\
 $|V_{cd}|$ & $0.22486\pm 0.00067$ \vspace{0.05cm}\\
 $|V_{cs}|$ & $0.97349\pm 0.00016$ \vspace{0.05cm}\\
 $|V_{cb}|$ & $0.04182\pm 0.00074$ \vspace{0.05cm}\\
 $|V_{td}|$ & $0.00857\pm 0.00018$ \vspace{0.05cm}\\
 $|V_{ts}|$ & $0.04110\pm 0.00072$ \vspace{0.05cm}\\
 $|V_{tb}|$ & $0.999118\pm 0.000031$ \vspace{0.05cm}\\
$\mathcal{J}$ & $(3.08 \pm 0.13) \cross 10^{-5}$ \vspace{0.05cm}\\
\hline 
\end{tabular} 
\caption{Norm of the numerical values of the matrix elements $V_{CKM}$ that were used in the adjustment of $\chi^2$, $\mathcal{J}$ is the Jarlskog invariant. Values can be found in \cite{Workman:2022ynf}.}
\label{tablamedidas}
\end{center}
\end{table}
Table \ref{tablamasas} contains the data on the masses that will be used for the theoretical calculation of the matrix elements of $V_{CKM}$ with the help of the equations in (\ref{elem:ckm_S3SM}) and that are present in $\chi^2$. Table \ref{tablamedidas} contains the experimental values of the terms appearing in the $\chi^2$ function. The masses in the third column are in the $\overline{MS}$ scheme at the $Z$ boson scale, $M_Z$. Masses were calculated with the Mathematica RunDec package \cite{chetyrkin2000rundec}. The free parameters were adjusted in such a way that $\chi^2$ is minimal.

The ratios of the masses are presented in Table \ref{tablamedidasratios}.  \begin{table}[ht]
\tabcolsep 32pt
\begin{center}
\begin{tabular}{cc}
\hline 
Ratios & Center value\rule[-2ex]{0pt}{5ex}\\
\hline \\
 $\widetilde{\sigma}_u$ & $7.032 \cross10^{-6}$ \vspace{0.05cm}\\
 $\widetilde{\sigma}_d$ & $9.44\cross10^{-4}$ \vspace{0.05cm}\\
 $\widetilde{\sigma}_s$ & $0.0190$   \\
 $\widetilde{\sigma}_c$  & $0.00379$ \\
\hline 
\end{tabular} 
\caption{Central values of the ratios of the masses.}
\label{tablamedidasratios}
\end{center}
\end{table}
The minimization was computed with Mathematica for the free parameters $K_{9u}$, $K_{9d}$, $\phi_{2u}$ and $\phi_{2d}$. It is worth noting that for setup 1, we have three parameters, which are insufficient to obtain an adequate CKM matrix. 
\\
For setup 2, we know that with 4 parameters we should be able to reproduce the $V_{CKM}$ matrix.  We actually did the fit using all entries of the matrix, since we will use these results to extract the possible range of values of the Lagrangian parameters.
Table \ref{tablaajustesf} shows the values of the fit parameters for the ratios of the fixed masses. In this setting, the mixing matrix is expressed as
\begin{align}
V_{CKM}^{th}=
\left(
\begin{array}{ccc}
0.97435 & 0.2250 & 0.00369 \\
0.22486 & 0.97349 & 0.04182 \\
0.00857 & 0.04110 & 0.999118
\end{array}
\right),
\end{align}
and the Jarlskog invariant $\mathcal{J}^{th}=3.07\cross10^{-5}$.\\

It is important to note that the value of the function $\chi^2$ is overfitted. This is because only 4 out of the nine parameters of the matrix need to be accurately adjusted, since the VCKM matrix is unitary. 

Now, using the values in Table \ref{tablaajustesf}, the expressions (\ref{eq:Csup}), and the values for $0.01 \lesssim \tan\theta \lesssim 3$, we can extract a range for the Lagrangian values that appear in the mass matrices. These values for $\tan\theta$ are taken from ref.~\cite{Espinoza:2018itz}, where an analysis of a non-modular \sy symmetry  is performed, in a 3 Higgs doublet model plus an inert one. The values are the ones that best comply with  the LHC bounds for extended Higgs sectors, assuming an alignment limit, where the SM Higgs boson is maximally coupled to the gauge bosons and the other scalars are decoupled. 
The results for the  Lagrangian parameters $C_2, C_4, C_5, C_9$, and $\beta$, given in Eq.~(\ref{eq:Csup}) are shown in figure \ref{Cud}, as a function of $\tan\theta$, for the up and down sector respectively. In this analysis the values from Table \ref{tablaajustesf}, $v=246.22~GeV$, and $y_1 =0.0754 $ (the latter calculated from Eqs.(\ref{formass3a-1}) and (\ref{formass3a})), were used. This way it is possible to extract the values of some of the Lagrangian parameters that best comply with experimental data  in terms of the scalar potential parameter $\tan\theta$.  Although this is a particular setup, it exemplifies well the importance of using 
the results of the minimization of the scalar potential and its viable parameter space, together with the evaluation of the Yukawa couplings at specific fixed modular points, in order to reduce the amount of free parameters in these kind of models, and to extract a range of viability for some of the Lagrangian parameters.  

  \begin{figure*}[t]
    \centering
    \includegraphics[width=0.3\linewidth]{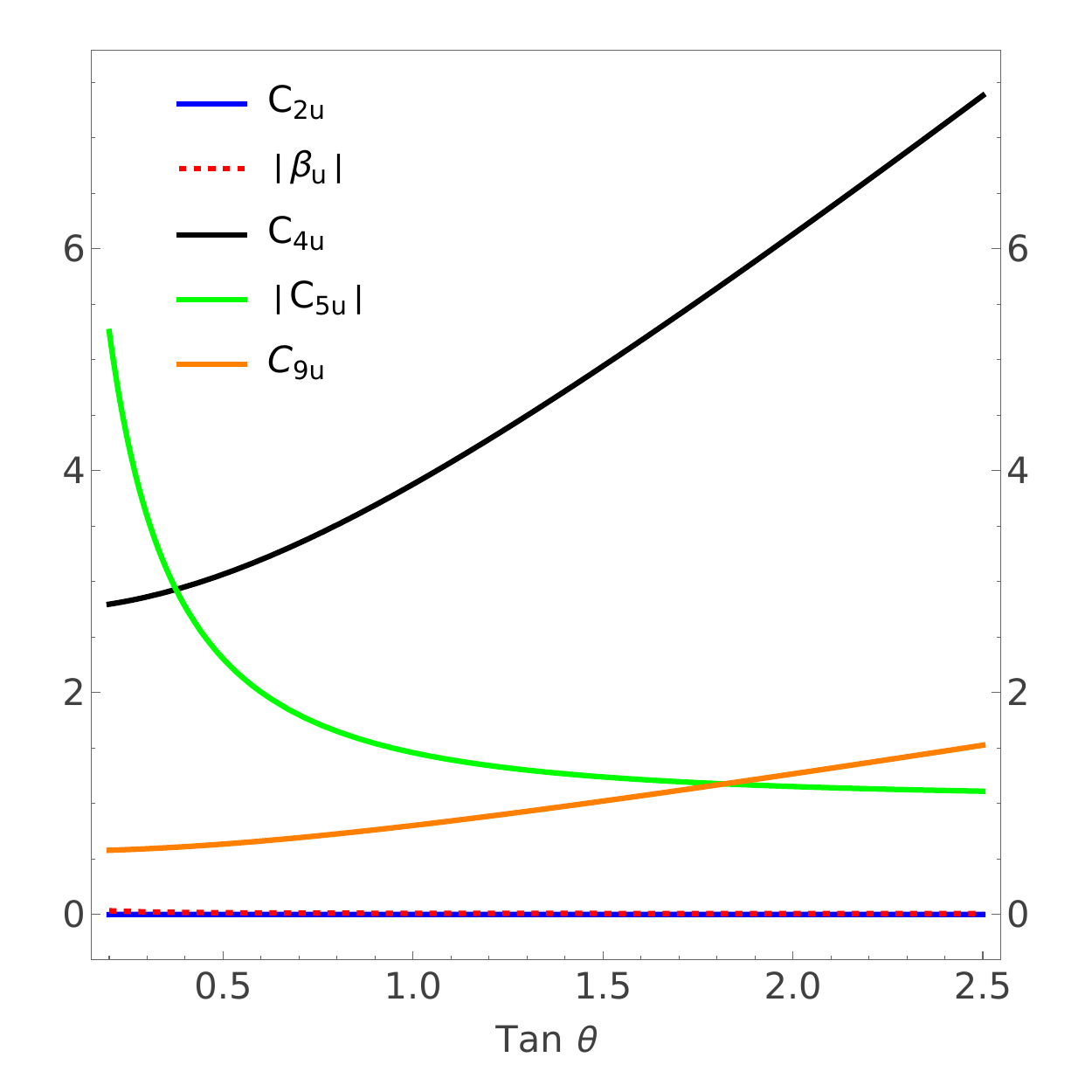}\hspace{3cm}
    \includegraphics[width=0.3\linewidth]{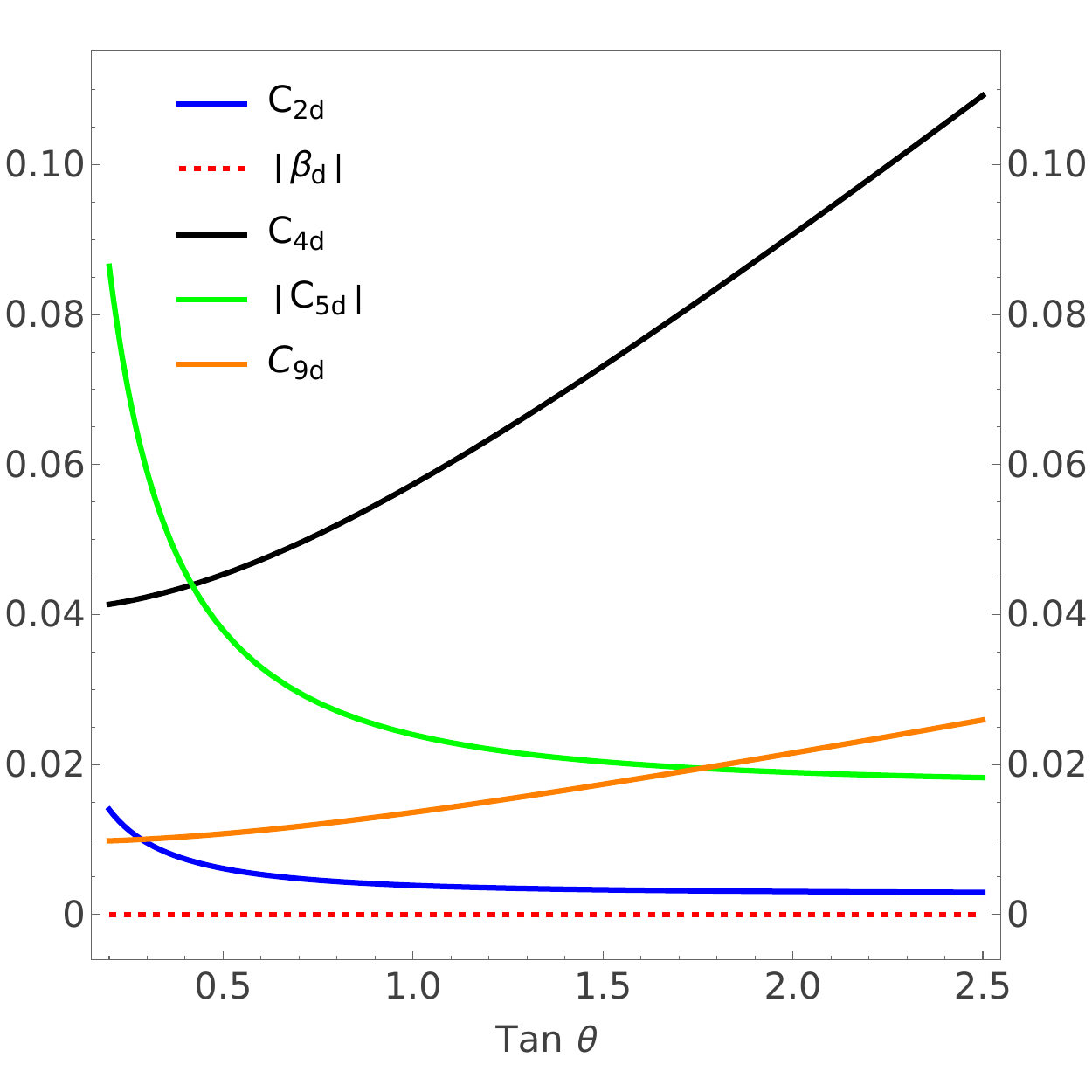}
    \caption{Values of the Lagrangian parameters as a function of $\tan\theta$ for up-type (left) and down-type (right). For complex parameters, absolute values are shown.}
    \label{Cud}
\end{figure*}

\begin{table}[ht]
\tabcolsep 30pt
\begin{center}
\begin{tabular}{cc}
\hline 
 Parameters & Values in the fit\rule[-2ex]{0pt}{5ex}\\
\hline \\
 $K_{9u}$ & $0.816393$ \vspace{0.05cm} \\
 $K_{9d}$ & $0.828604$ \vspace{0.05cm}\\
 $\phi_{1u}$ & $1.63797$ \vspace{0.05cm}\\
 $\phi_{1d}$  & $0$ \vspace{0.05cm}\\
 $\phi_{2u}$ & $0.0981477$ \vspace{0.05cm}\\
 $\phi_{2d}$  & $0$ \vspace{0.05cm}\\
 $\chi^2$ & $0.00070$ \vspace{0.05cm}\\
\hline  
\end{tabular} 
\caption{Values of the free parameters for the adjustment with the values of the ratios of the masses fixed in their central value and the respective obtaining of $\chi^2$.}
\label{tablaajustesf}
\end{center}
\end{table}

\section{Conclusions}

Starting from modular symmetry and the finite modular group $\Gamma(2)$, along with three Higgs doublets, an extension of the Standard Model has been successfully constructed. The model was developed through the introduction of modular forms. Under the proposed symmetry, and after assigning appropriate modular weights and charges to the fields, the Yukawa sector Lagrangian was derived. The inclusion of the Higgs potential was achieved by imposing modular weight constraints on both the couplings and the Higgs doublets.

The freedom in parameter assignments—both in the $\Gamma(2)$ symmetry and under modular invariance—is considerable, as the introduction of an additional doublet of modular forms generates more product combinations compared to a conventional $S_3$ symmetry. After spontaneous symmetry breaking, the quark mass matrices were obtained, and the total number of free parameters was found to be 47 real parameters (21 complex, 5 real). The increased number of parameters, compared with other models, arises from the presence of three Higgs doublets, and the modular forms} which introduce additional terms in the Lagrangian. However, evaluating the Yukawa couplings at the fixed point $\tau=i$, and using previous results on the minimization of the Higgs potential it is possible to reduce significantly the number of free parameters.

To construct the two texture zero mass matrix, two setups were proposed, in which some parameters were fixed for simplicity. Analytical expressions were used to compute the mixing matrix. The first setup, with three effective parameters proved inadequate for obtaining a realistic $V_{\text{CKM}}$ matrix, as its parameters were overly constrained. The second setup, which includes one additional free parameter, allowed the construction of a suitable $V_{\text{CKM}}$ matrix since its parameters are less restricted.

It is noteworthy that under modular $S_3$,  and after substituting the relationship among vev's from the minimization of the scalar potential, the $V_{\text{CKM}}$ matrix exhibits no zero entries, unlike some models employing conventional $S_3$ symmetry and three Higgs doublets. The total number of effective free parameters in the up- and down-type quark mass matrices in the setup 2 was four. These parameters were fitted to the experimental observables of $V_{\text{CKM}}$ and the Jarlskog invariant, yielding the best fit with $\chi^2 = 0.0070$. As previously mentioned, this result indicates a certain degree of overfitting, since the number of free parameters equals the minimum required to construct a $3\times3$ unitary matrix. The unitarity of the $V_{\text{CKM}}$ matrix was verified, and its construction follows the procedure described in \cite{canales2013quark,barranco2010universal}.

Therefore, assuming a \sy modular symmetry and employing modular forms with the indicated parameter assignments, together with evaluation at the modular fixed point $i$, allows  for the construction of viable texture zero mass matrices, suggesting that the  symmetry provides a good approximation to the observed quark mixing pattern. Moreover, using previous results from the scalar potential and its phenomenologically allowed parameter space, it is possible to extract values for the Lagrangian parameters, which could provide with  further information for constructing viable models.

An extension of this work could explore alternative field assignments under $S_3$ and different modular weight configurations. The inclusion of Higgs potentials whose couplings transform as modular forms could offer additional flexibility in constructing mass matrices or reduce the number of free parameters, potentially achieving a closer fit to the observed quark mixing parameters. Furthermore, hybrid models involving other finite modular groups $\Gamma_N$, such as $A_4$ (isomorphic to $\Gamma(3)$), could also be considered for future developments.

\section*{Acknowledgements}
We acknowledge useful discussions with M.C. Chun, M. Gómez-Bock, A. Pérez-Ramírez, S. Petcov, M. Ratz, and S. Ramos-Sánchez.  This work is partly supported by UNAM project PAPIIT  IN111224 and SECIHTI project CBF2023-2024-548. MM is grateful to the Mainz Institute for Theoretical Physics
(MITP) of the Cluster of Excellence PRISMA+ (Project ID 390831469), for its hospitality and partial support during the development of this work. 

\bibliographystyle{unsrtnat} 
\bibliography{references}

\begin{thebibliography}{63}
\providecommand{\natexlab}[1]{#1}
\providecommand{\url}[1]{\texttt{#1}}
\expandafter\ifx\csname urlstyle\endcsname\relax
  \providecommand{\doi}[1]{doi: #1}\else
  \providecommand{\doi}{doi: \begingroup \urlstyle{rm}\Url}\fi

\bibitem[{Ishimori, Hajime and Kobayashi, Tatsuo and Ohki, Hiroshi and Shimizu,
  Yusuke and Okada, Hiroshi and Tanimoto, Morimitsu}(2010)]{Ishimori:2010au}
{Ishimori, Hajime and Kobayashi, Tatsuo and Ohki, Hiroshi and Shimizu, Yusuke
  and Okada, Hiroshi and Tanimoto, Morimitsu}.
\newblock {Non-Abelian Discrete Symmetries in Particle Physics}.
\newblock \emph{Prog. Theor. Phys. Suppl.}, 183:\penalty0 1--163, 2010.
\newblock \doi{10.1143/PTPS.183.1}.

\bibitem[Lam(2001)]{lam20012}
CS~Lam.
\newblock A 2-3 symmetry in neutrino oscillations.
\newblock \emph{Physics Letters B}, 507\penalty0 (1-4):\penalty0 214--218,
  2001.
\newblock \doi{10.1016/S0370-2693(01)00465-8}.

\bibitem[Dixon et~al.(1987)Dixon, Friedan, Martinec, and
  Shenker]{dixon1987conformal}
Lance Dixon, Daniel Friedan, Emil Martinec, and Stephen Shenker.
\newblock The conformal field theory of orbifolds.
\newblock \emph{Nuclear Physics B}, 282:\penalty0 13--73, 1987.
\newblock \doi{10.1016/0550-3213(87)90676-6}.
\newblock URL \url{https://doi.org/10.1016/0550-3213(87)90676-6}.

\bibitem[Altarelli and Feruglio(2006)]{Altarelli:2005yx}
Guido Altarelli and Ferruccio Feruglio.
\newblock {Tri-bimaximal neutrino mixing, A(4) and the modular symmetry}.
\newblock \emph{Nucl. Phys. B}, 741:\penalty0 215--235, 2006.
\newblock \doi{10.1016/j.nuclphysb.2006.02.015}.

\bibitem[Chen et~al.(2020{\natexlab{a}})Chen, Ramos-Sánchez, and
  Ratz]{CHEN2020135153}
Mu-Chun Chen, Saúl Ramos-Sánchez, and Michael Ratz.
\newblock A note on the predictions of models with modular flavor symmetries.
\newblock \emph{Physics Letters B}, 801:\penalty0 135153, 2020{\natexlab{a}}.
\newblock ISSN 0370-2693.
\newblock \doi{https://doi.org/10.1016/j.physletb.2019.135153}.

\bibitem[Feruglio(2019{\natexlab{a}})]{Feruglio:2017spp}
Ferruccio Feruglio.
\newblock \emph{{Are neutrino masses modular forms?}}, pages 227--266.
\newblock 2019{\natexlab{a}}.
\newblock \doi{10.1142/9789813238053_0012}.

\bibitem[Meloni and Parriciatu(2023)]{Meloni2023}
Davide Meloni and Matteo Parriciatu.
\newblock A simplest modular s3 model for leptons.
\newblock \emph{Journal of High Energy Physics}, 2023\penalty0 (9):\penalty0
  43, Sep 2023.
\newblock ISSN 1029-8479.
\newblock \doi{10.1007/JHEP09(2023)043}.
\newblock URL \url{https://doi.org/10.1007/JHEP09(2023)043}.

\bibitem[Penedo and Petcov(2024)]{Penedo:2024gtb}
J.~T. Penedo and S.~T. Petcov.
\newblock {Finite modular symmetries and the strong CP problem}.
\newblock \emph{INSPIRE HEP}, 4 2024.

\bibitem[Feruglio et~al.(2024)Feruglio, Parriciatu, Strumia, and
  Titov]{Feruglio:2024ytl}
Ferruccio Feruglio, Matteo Parriciatu, Alessandro Strumia, and Arsenii Titov.
\newblock {Solving the strong CP problem without axions}.
\newblock \emph{Arxiv}, 6 2024.

\bibitem[Higaki et~al.(2024)Higaki, Kobayashi, Nasu, and
  Otsuka]{Higaki:2024pql}
Tetsutaro Higaki, Tatsuo Kobayashi, Kaito Nasu, and Hajime Otsuka.
\newblock {Spontaneous CP violation and partially broken modular flavor
  symmetries}.
\newblock \emph{INSPIRE HEP}, 5 2024.

\bibitem[Jung and Kawamura(2024)]{Jung:2024bgi}
Tae~Hyun Jung and Junichiro Kawamura.
\newblock Finite modular majoron.
\newblock \emph{Journal of High Energy Physics}, 2024\penalty0 (7):\penalty0
  145, Jul 2024.

\bibitem[Kobayashi et~al.(2025)Kobayashi, Okada, and
  Orikasa]{Kobayashi:2025hnc}
Tatsuo Kobayashi, Hiroshi Okada, and Yuta Orikasa.
\newblock {Zee-Babu model in a non-holomorphic modular $A_4$ symmetry and
  modular stabilization}.
\newblock 2 2025.

\bibitem[Diamond and Shurman(2005)]{diamond2005first}
Fred Diamond and Jerry~Michael Shurman.
\newblock \emph{A first course in modular forms}, volume 228.
\newblock Springer, 2005.

\bibitem[Qu and Ding(2024{\natexlab{a}})]{Qu:2024rns}
Bu-Yao Qu and Gui-Jun Ding.
\newblock {Non-holomorphic modular flavor symmetry}.
\newblock \emph{JHEP}, 08:\penalty0 136, 2024{\natexlab{a}}.
\newblock \doi{10.1007/JHEP08(2024)136}.

\bibitem[Chen et~al.(2022)Chen, Knapp–Pérez, Ramos–Hamud,
  Ramos–Sánchez, Ratz, and Shukla]{CHEN2022136843}
Mu–Chun Chen, Víctor Knapp–Pérez, Mario Ramos–Hamud, Saúl
  Ramos–Sánchez, Michael Ratz, and Shreya Shukla.
\newblock Quasi–eclectic modular flavor symmetries.
\newblock \emph{Physics Letters B}, 824:\penalty0 136843, 2022.
\newblock ISSN 0370-2693.
\newblock \doi{https://doi.org/10.1016/j.physletb.2021.136843}.

\bibitem[Abel et~al.(2015)Abel, Dienes, and Mavroudi]{PhysRevD.91.126014}
Steven Abel, Keith~R. Dienes, and Eirini Mavroudi.
\newblock Towards a nonsupersymmetric string phenomenology.
\newblock \emph{Phys. Rev. D}, 91:\penalty0 126014, Jun 2015.
\newblock \doi{10.1103/PhysRevD.91.126014}.
\newblock URL \url{https://link.aps.org/doi/10.1103/PhysRevD.91.126014}.

\bibitem[Almumin et~al.(2023)Almumin, Chen, Cheng, Knapp-Pérez, Li, Mondol,
  Ramos-Sánchez, Ratz, and Shukla]{universe9120512}
Yahya Almumin, Mu-Chun Chen, Murong Cheng, Víctor Knapp-Pérez, Yulun Li,
  Adreja Mondol, Saúl Ramos-Sánchez, Michael Ratz, and Shreya Shukla.
\newblock Neutrino flavor model building and the origins of flavor and cp
  violation.
\newblock \emph{Universe}, 9\penalty0 (12), 2023.
\newblock ISSN 2218-1997.
\newblock \doi{10.3390/universe9120512}.
\newblock URL \url{https://www.mdpi.com/2218-1997/9/12/512}.

\bibitem[Ratz(2024)]{Ratz:2024imd}
Michael Ratz.
\newblock {Aspects of Modular Flavor Symmetries}.
\newblock In \emph{{2024 Moriond EW}}, pages~--, 5 2024.

\bibitem[Okada and Orikasa(2025)]{Okada:2025jjo}
Hiroshi Okada and Yuta Orikasa.
\newblock {A radiative seesaw in a non-holomorphic modular $S_3$ flavor
  symmetry}.
\newblock 1 2025.

\bibitem[Ding et~al.(2025)Ding, Lu, Petcov, and Qu]{Ding2025}
Gui-Jun Ding, Jun-Nan Lu, S.~T. Petcov, and Bu-Yao Qu.
\newblock Non-holomorphic modular s4 lepton flavour models.
\newblock \emph{Journal of High Energy Physics}, 2025\penalty0 (1):\penalty0
  191, Jan 2025.
\newblock ISSN 1029-8479.
\newblock \doi{10.1007/JHEP01(2025)191}.
\newblock URL \url{https://doi.org/10.1007/JHEP01(2025)191}.

\bibitem[Qu and Ding(2024{\natexlab{b}})]{Qu2024}
Bu-Yao Qu and Gui-Jun Ding.
\newblock Non-holomorphic modular flavor symmetry.
\newblock \emph{Journal of High Energy Physics}, 2024\penalty0 (8):\penalty0
  136, Aug 2024{\natexlab{b}}.
\newblock ISSN 1029-8479.
\newblock \doi{10.1007/JHEP08(2024)136}.
\newblock URL \url{https://doi.org/10.1007/JHEP08(2024)136}.

\bibitem[Nomura and Okada(2019)]{NOMURA2019134799}
Takaaki Nomura and Hiroshi Okada.
\newblock A modular a4 symmetric model of dark matter and neutrino.
\newblock \emph{Physics Letters B}, 797:\penalty0 134799, 2019.
\newblock ISSN 0370-2693.
\newblock \doi{https://doi.org/10.1016/j.physletb.2019.134799}.
\newblock URL
  \url{https://www.sciencedirect.com/science/article/pii/S0370269319305039}.

\bibitem[Qu et~al.(2025)Qu, Lu, and Ding]{Qu:2025ddz}
Bu-Yao Qu, Jun-Nan Lu, and Gui-Jun Ding.
\newblock {Non-holomorphic modular flavor symmetry and odd weight polyharmonic
  Maa{\ss} form}.
\newblock \emph{JHEP}, 11:\penalty0 140, 2025.
\newblock \doi{10.1007/JHEP11(2025)140}.

\bibitem[Li et~al.(2024)Li, Lu, and Ding]{Li:2024svh}
Cai-Chang Li, Jun-Nan Lu, and Gui-Jun Ding.
\newblock {Non-holomorphic modular A$_{5}$ symmetry for lepton masses and
  mixing}.
\newblock \emph{JHEP}, 12:\penalty0 189, 2024.
\newblock \doi{10.1007/JHEP12(2024)189}.

\bibitem[Kobayashi et~al.(2019)Kobayashi, Shimizu, Takagi, Tanimoto, Tatsuishi,
  and Uchida]{kobayashi2019finite}
Tatsuo Kobayashi, Yusuke Shimizu, Kenta Takagi, Morimitsu Tanimoto, Takuya~H
  Tatsuishi, and Hikaru Uchida.
\newblock Finite modular subgroups for fermion mass matrices and baryon/lepton
  number violation.
\newblock \emph{Physics Letters B}, 794:\penalty0 114--121, 2019.
\newblock \doi{10.1016/j.physletb.2019.05.034}.

\bibitem[Feruglio(2019{\natexlab{b}})]{feruglio2019neutrino}
Ferruccio Feruglio.
\newblock Are neutrino masses modular forms?
\newblock \emph{arXiv preprint arXiv:1706.08749}, -, 2019{\natexlab{b}}.
\newblock \doi{10.1142/9789813238053\_0012}.

\bibitem[Kobayashi et~al.(2018{\natexlab{a}})Kobayashi, Tanaka, and
  Tatsuishi]{PhysRevD.98.016004}
Tatsuo Kobayashi, Kentaro Tanaka, and Takuya~H. Tatsuishi.
\newblock Neutrino mixing from finite modular groups.
\newblock \emph{Phys. Rev. D}, 98:\penalty0 016004, Jul 2018{\natexlab{a}}.
\newblock \doi{10.1103/PhysRevD.98.016004}.
\newblock URL \url{https://link.aps.org/doi/10.1103/PhysRevD.98.016004}.

\bibitem[Penedo and Petcov(2019)]{penedo2019lepton}
JT~Penedo and ST~Petcov.
\newblock Lepton masses and mixing from modular s4 symmetry.
\newblock \emph{Nuclear Physics B}, 939:\penalty0 292--307, 2019.
\newblock \doi{doi.org/10.1016/j.nuclphysb.2018.12.016}.
\newblock URL \url{https://doi.org/10.1016/j.nuclphysb.2018.12.016}.

\bibitem[Novichkov et~al.(2019)Novichkov, Penedo, Petcov, and
  Titov]{novichkov2019modular}
PP~Novichkov, JT~Penedo, ST~Petcov, and AV~Titov.
\newblock Modular a5 symmetry for flavour model building.
\newblock \emph{Journal of High Energy Physics}, 2019\penalty0 (4):\penalty0
  174, 2019.
\newblock \doi{10.1007/JHEP04(2019)174}.
\newblock URL \url{https://doi.org/10.1007/JHEP04(2019)174}.

\bibitem[de~Medeiros~Varzielas et~al.(2023)de~Medeiros~Varzielas, Levy, Penedo,
  and Petcov]{deMedeirosVarzielas2023}
I.~de~Medeiros~Varzielas, M.~Levy, J.~T. Penedo, and S.~T. Petcov.
\newblock Quarks at the modular s4 cusp.
\newblock \emph{Journal of High Energy Physics}, 2023\penalty0 (9):\penalty0
  196, Sep 2023.
\newblock ISSN 1029-8479.
\newblock \doi{10.1007/JHEP09(2023)196}.
\newblock URL \url{https://doi.org/10.1007/JHEP09(2023)196}.

\bibitem[Kubo et~al.(2003)Kubo, Mondragon, Mondragon, and
  Rodriguez-Jauregui]{Kubo:2003iw}
J.~Kubo, A.~Mondragon, M.~Mondragon, and E.~Rodriguez-Jauregui.
\newblock {The Flavor symmetry}.
\newblock \emph{Prog. Theor. Phys.}, 109:\penalty0 795--807, 2003.
\newblock \doi{10.1143/PTP.109.795}.
\newblock [Erratum: Prog. Theor. Phys.114,287(2005)].

\bibitem[Canales et~al.(2013)Canales, Mondragón, Mondragón, Salazar, and
  Velasco-Sevilla]{canales2013quark}
F~González Canales, A~Mondragón, M~Mondragón, UJ~Salda{\~n}a Salazar, and
  L~Velasco-Sevilla.
\newblock Quark sector of s 3 models: classification and comparison with
  experimental data.
\newblock \emph{Physical Review D}, 88\penalty0 (9):\penalty0 096004, 2013.
\newblock \doi{10.1103/PhysRevD.88.096004}.

\bibitem[Mondragon et~al.(2007)Mondragon, Mondragon, and
  Peinado]{Mondragon:2007af}
A.~Mondragon, M.~Mondragon, and E.~Peinado.
\newblock {Lepton masses, mixings and FCNC in a minimal S(3)-invariant
  extension of the Standard Model}.
\newblock \emph{Phys. Rev.}, D76:\penalty0 076003, 2007.
\newblock \doi{10.1103/PhysRevD.76.076003}.

\bibitem[Das and Dey(2014)]{Das:2014fea}
Dipankar Das and Ujjal~Kumar Dey.
\newblock {Analysis of an extended scalar sector with $S_3$ symmetry}.
\newblock \emph{Phys. Rev.}, D89\penalty0 (9):\penalty0 095025, 2014.
\newblock \doi{10.1103/PhysRevD.91.039905, 10.1103/PhysRevD.89.095025}.
\newblock [Erratum: Phys. Rev.D91,no.3,039905(2015)].

\bibitem[G\'omez-Bock et~al.(2021)G\'omez-Bock, Mondrag\'on, and
  P\'erez-Mart\'\i{}nez]{Gomez-Bock:2021uyu}
M.~G\'omez-Bock, M.~Mondrag\'on, and A.~P\'erez-Mart\'\i{}nez.
\newblock {Scalar and gauge sectors in the 3-Higgs Doublet Model under the
  $S_3$ symmetry}.
\newblock \emph{Eur. Phys. J. C}, 81\penalty0 (10):\penalty0 942, 2021.
\newblock \doi{10.1140/epjc/s10052-021-09731-3}.

\bibitem[Petcov and Tanimoto(2023{\natexlab{a}})]{Petcov20231}
S.~T. Petcov and M.~Tanimoto.
\newblock A4 modular flavour model of quark mass hierarchies close to the fixed
  point $\tau = i\infty$.
\newblock \emph{Journal of High Energy Physics}, 2023\penalty0 (8):\penalty0
  86, Aug 2023{\natexlab{a}}.
\newblock ISSN 1029-8479.
\newblock \doi{10.1007/JHEP08(2023)086}.
\newblock URL \url{https://doi.org/10.1007/JHEP08(2023)086}.

\bibitem[Petcov and Tanimoto(2023{\natexlab{b}})]{Petcov20232}
S.~T. Petcov and M.~Tanimoto.
\newblock $a_4$ modular flavour model of quark mass hierarchies close to the
  fixed point $\tau = \omega$.
\newblock \emph{The European Physical Journal C}, 83\penalty0 (7):\penalty0
  579, Jul 2023{\natexlab{b}}.
\newblock ISSN 1434-6052.
\newblock \doi{10.1140/epjc/s10052-023-11727-0}.
\newblock URL \url{https://doi.org/10.1140/epjc/s10052-023-11727-0}.

\bibitem[Cremades et~al.(2004)Cremades, Ibá{\c{n}}ez, and
  Marchesano]{cremades2004computing}
Daniel Cremades, Luis~E Ibá{\c{n}}ez, and Fernando Marchesano.
\newblock Computing yukawa couplings from magnetized extra dimensions.
\newblock \emph{Journal of High Energy Physics}, 2004\penalty0 (05):\penalty0
  079, 2004.
\newblock \doi{10.1088/1126-6708/2004/05/079}.

\bibitem[Kobayashi et~al.(2018{\natexlab{b}})Kobayashi, Nagamoto, Takada,
  Tamba, and Tatsuishi]{kobayashi2018modularstring}
Tatsuo Kobayashi, Satoshi Nagamoto, Shintaro Takada, Shio Tamba, and Takuya~H
  Tatsuishi.
\newblock Modular symmetry and non-abelian discrete flavor symmetries in string
  compactification.
\newblock \emph{Physical Review D}, 97\penalty0 (11):\penalty0 116002,
  2018{\natexlab{b}}.
\newblock \doi{10.1103/PhysRevD.97.116002}.

\bibitem[Spaliński(1992)]{spalinski1992duality}
Micha{\l} Spaliński.
\newblock Duality transformations in twisted narain compactifications.
\newblock \emph{Nuclear Physics B}, 377\penalty0 (1-2):\penalty0 339--368,
  1992.
\newblock \doi{10.1016/0550-3213(92)90027-9}.

\bibitem[Hamidi and Vafa(1987)]{hamidi1987interactions}
Shahram Hamidi and Cumrun Vafa.
\newblock Interactions on orbifolds.
\newblock \emph{Nuclear Physics B}, 279\penalty0 (3-4):\penalty0 465--513,
  1987.
\newblock \doi{10.1016/0550-3213(87)90006-X}.

\bibitem[Erler et~al.(1992)Erler, Jungnickel, and Lauer]{erler1992dependence}
J~Erler, D~Jungnickel, and J~Lauer.
\newblock Dependence of yukawa couplings on the axionic background moduli of z
  n orbifolds.
\newblock \emph{Physical Review D}, 45\penalty0 (10):\penalty0 3651, 1992.
\newblock \doi{10.1103/PhysRevD.45.3651}.

\bibitem[Lauer et~al.(1989)Lauer, Mas, and Nilles]{lauer1989duality}
J~Lauer, J~Mas, and Hans~Peter Nilles.
\newblock Duality and the role of nonperturbative effects on the world-sheet.
\newblock \emph{Physics Letters B}, 226\penalty0 (3-4):\penalty0 251--256,
  1989.
\newblock \doi{10.1016/0370-2693(89)91190-8}.

\bibitem[Nomura et~al.(2020)Nomura, Okada, and Popov]{nomura2020modular}
Takaaki Nomura, Hiroshi Okada, and Oleg Popov.
\newblock A modular a4 symmetric scotogenic model.
\newblock \emph{Physics Letters B}, 803:\penalty0 135294, 2020.

\bibitem[Nomura et~al.(2021)Nomura, Okada, and Patra]{nomura2021inverse}
Takaaki Nomura, Hiroshi Okada, and Sudhanwa Patra.
\newblock An inverse seesaw model with a4-modular symmetry.
\newblock \emph{Nuclear Physics B}, 967:\penalty0 115395, 2021.

\bibitem[Nomura and Okada(2021)]{nomura2021two}
Takaaki Nomura and Hiroshi Okada.
\newblock A two loop induced neutrino mass model with modular a4 symmetry.
\newblock \emph{Nuclear Physics B}, 966:\penalty0 115372, 2021.

\bibitem[Chen et~al.(2020{\natexlab{b}})Chen, Ramos-Sánchez, and
  Ratz]{chen2020note}
Mu-Chun Chen, Saúl Ramos-Sánchez, and Michael Ratz.
\newblock A note on the predictions of models with modular flavor symmetries.
\newblock \emph{Physics Letters B}, 801:\penalty0 135153, 2020{\natexlab{b}}.
\newblock \doi{10.1016/j.physletb.2019.135153}.

\bibitem[Novichkov et~al.(2021{\natexlab{a}})Novichkov, Penedo, and
  Petcov]{novichkov2021fermion}
PP~Novichkov, JT~Penedo, and ST~Petcov.
\newblock Fermion mass hierarchies, large lepton mixing and residual modular
  symmetries.
\newblock \emph{Journal of High Energy Physics}, 2021\penalty0 (4):\penalty0
  1--49, 2021{\natexlab{a}}.
\newblock \doi{10.1007/JHEP04(2021)206}.

\bibitem[Novichkov et~al.(2021{\natexlab{b}})Novichkov, Penedo, and
  Petcov]{novichkov2021double}
PP~Novichkov, JT~Penedo, and ST~Petcov.
\newblock Double cover of modular s4 for flavour model building.
\newblock \emph{Nuclear Physics B}, 963:\penalty0 115301, 2021{\natexlab{b}}.
\newblock \doi{10.1016/j.nuclphysb.2020.115301}.

\bibitem[Harrison and Scott(2003)]{harrison2003permutation}
PF~Harrison and WG~Scott.
\newblock Permutation symmetry, tri-bimaximal neutrino mixing and the s3 group
  characters.
\newblock \emph{Physics Letters B}, 557\penalty0 (1-2):\penalty0 76--86, 2003.
\newblock \doi{10.1016/S0370-2693(03)00183-7}.

\bibitem[Liu and Ding(2019)]{liu2019neutrino}
Xiang-Gan Liu and Gui-Jun Ding.
\newblock Neutrino masses and mixing from double covering of finite modular
  groups.
\newblock \emph{Journal of High Energy Physics}, 2019\penalty0 (8):\penalty0
  1--21, 2019.
\newblock \doi{10.1007/JHEP08(2019)134}.

\bibitem[Schoeneberg(2012)]{schoeneberg2012elliptic}
Bruno Schoeneberg.
\newblock \emph{Elliptic modular functions: an introduction}, volume 203.
\newblock Springer Science \& Business Media, 2012.

\bibitem[Gunning(2016)]{gunning2016lectures}
Robert~C Gunning.
\newblock \emph{Lectures on Modular Forms.(AM-48), Volume 48}.
\newblock Princeton University Press, 2016.

\bibitem[Katok(1992)]{katok1992fuchsian}
Svetlana Katok.
\newblock \emph{Fuchsian groups}.
\newblock University of Chicago press, 1992.

\bibitem[de~Adelhart~Toorop et~al.(2012)de~Adelhart~Toorop, Feruglio, and
  Hagedorn]{deAdelhartToorop:2011re}
Reinier de~Adelhart~Toorop, Ferruccio Feruglio, and Claudia Hagedorn.
\newblock {Finite Modular Groups and Lepton Mixing}.
\newblock \emph{Nucl. Phys. B}, 858:\penalty0 437--467, 2012.
\newblock \doi{10.1016/j.nuclphysb.2012.01.017}.

\bibitem[Beltran et~al.(2009)Beltran, Mondragon, and
  Rodriguez-Jauregui]{Beltran:2009zz}
O.~Felix Beltran, M.~Mondragon, and E.~Rodriguez-Jauregui.
\newblock {Conditions for vacuum stability in an S(3) extension of the standard
  model}.
\newblock \emph{J. Phys. Conf. Ser.}, 171:\penalty0 012028, 2009.
\newblock \doi{10.1088/1742-6596/171/1/012028}.

\bibitem[Barradas-Guevara et~al.(2014)Barradas-Guevara, Félix-Beltrán, and
  Rodríguez-Jáuregui]{Barradas-Guevara:2014yoa}
E.~Barradas-Guevara, O.~Félix-Beltrán, and E.~Rodríguez-Jáuregui.
\newblock {Trilinear self-couplings in an S(3) flavored Higgs model}.
\newblock \emph{Phys. Rev.}, D90\penalty0 (9):\penalty0 095001, 2014.
\newblock \doi{10.1103/PhysRevD.90.095001}.

\bibitem[Fritzsch and Xing(2000)]{FRITZSCH20001}
H~Fritzsch and Z.-Z Xing.
\newblock Mass and flavor mixing schemes of quarks and leptons.
\newblock \emph{Progress in Particle and Nuclear Physics}, 45\penalty0
  (1):\penalty0 1--81, 2000.
\newblock ISSN 0146-6410.
\newblock \doi{https://doi.org/10.1016/S0146-6410(00)00102-2}.

\bibitem[Xing(2020)]{Xing:2020ijf}
Zhi-zhong Xing.
\newblock {Flavor structures of charged fermions and massive neutrinos}.
\newblock \emph{Phys. Rept.}, 854:\penalty0 1--147, 2020.
\newblock \doi{10.1016/j.physrep.2020.02.001}.

\bibitem[Barranco et~al.(2010)Barranco, Canales, and
  Mondragon]{barranco2010universal}
J~Barranco, F~Gonzalez Canales, and A~Mondragon.
\newblock Universal mass texture, cp violation and quark-lepton
  complementarity.
\newblock \emph{arXiv preprint arXiv:1004.3781}, -, 2010.
\newblock \doi{10.1088/1742-6596/171/1/012063}.

\bibitem[Workman et~al.(2022)]{Workman:2022ynf}
R.~L. Workman et~al.
\newblock {Review of Particle Physics}.
\newblock \emph{PTEP}, 2022:\penalty0 083C01, 2022.
\newblock \doi{10.1093/ptep/ptac097}.

\bibitem[Chetyrkin et~al.(2000)Chetyrkin, Kühn, and
  Steinhauser]{chetyrkin2000rundec}
KG~Chetyrkin, Johann~H Kühn, and Matthias Steinhauser.
\newblock Rundec: A mathematica package for running and decoupling of the
  strong coupling and quark masses.
\newblock \emph{Computer Physics Communications}, 133\penalty0 (1):\penalty0
  43--65, 2000.
\newblock \doi{10.1016/S0010-4655(00)00155-7}.

\bibitem[Espinoza et~al.(2019)Espinoza, Garcés, Mondragón, and
  Reyes-González]{Espinoza:2018itz}
C.~Espinoza, E.~A. Garcés, M.~Mondragón, and H.~Reyes-González.
\newblock {The $S3$ Symmetric Model with a Dark Scalar}.
\newblock \emph{Phys. Lett.}, B788:\penalty0 185--191, 2019.
\newblock \doi{10.1016/j.physletb.2018.11.028}.

\end{thebibliography}

\appendix
 \section{Construction of modular forms for \sy} \label{apendiceformas}
It is necessary to build the modular forms for the $\mathbf{2}$ representation of $S_3$. The procedure for its construction is based on the methods used in \cite{feruglio2019neutrino} by Feruglio and extended to $S_3$ in \cite{kobayashi2019finite} by Kobayashi.\\
First consider the transformation under $\Gamma$ of the expression $\frac{d}{d\tau}\log(\sum_i f_i(\tau))$, which is a modular form of weight 2. Using the transformation
\begin{equation}
    \frac{d}{d\tau} \rightarrow (c\tau+d)^2\frac{d}{d\tau},
\end{equation}
we have
\begin{align}
\frac{d}{d\tau}\log(\sum_i f_i(\tau)) &\rightarrow \sum_i (c\tau +d)k_i c \notag\\
&+\sum_i(c\tau+d)^2\frac{d}{d\tau}\log f_i(\tau). 
\end{align}
Thus, if $\sum_ik_i =0$, the derivative of the logarithm of $f_i$ is a modular form with weight 2.\\
A candidate modular form of weight 2 can be constructed from the Dedekind Eta functions, which are defined as
\begin{equation}
\eta(\tau) = q^{1/24} \prod_{n =1}^\infty (1-q^n),
\end{equation}
where $q = e^{2 \pi i \tau}$. The function $\eta(\tau)$ is a special type of modular form due to its modular weight of $1/2$ and is useful for constructing other modular forms since, under transformations of the generators $S$ and $T$ , a closed algebra is obtained. The transformations are

Under T
\begin{align}
&\eta(2\tau) \rightarrow e^{i \pi/6} \eta(2\tau), \nonumber \\
&\eta(\tau/2) \rightarrow \eta((\tau +1)/2),  \\
&\eta((\tau + 1)/2) \rightarrow e^{i \pi /12}\eta(\tau/2).  \nonumber 
\end{align}
and under S
\begin{align}
&\eta(2\tau) \rightarrow \sqrt{\frac{-i\tau}{2}} \eta(\tau/2), \nonumber \\
&\eta(\tau/2) \rightarrow \sqrt{-i3\tau} \eta(2\tau),  \\
&\eta((\tau + 1)/2) \rightarrow e^{-i\pi/12} \sqrt{-i\tau}\eta((\tau +1)/2). \nonumber 
\end{align}
The general modular form in terms of the function $\eta(\tau)$ can be written as
\begin{align}
Y(a,b,c| \tau) =\frac{d}{d\tau}\left( a\log\eta\left( \frac{\tau}{2}\right)+b\log\eta\left(\frac{\tau+1}{2}\right)+c\log\eta\left( 2\tau\right) \right), \label{general2}
\end{align}
that satisfies
\begin{eqnarray}
&a+b+c=0 \notag \\
&Y(a,b,c | \tau)\xrightarrow{S}\tau^2 Y(c,b,a | \tau)\\
&Y(a,b,c | \tau)\xrightarrow{T} Y(b,a,c | \tau) \notag
\end{eqnarray}
since
\begin{eqnarray}
\frac{d}{d\tau} \xrightarrow{S} \tau^2\frac{d}{d\tau} \hspace{1cm}\text{and}\hspace{1cm} \frac{d}{d\tau} \xrightarrow{T} \frac{d}{d\tau}.
\end{eqnarray}
Since $\Gamma_2\simeq S_3$, it is necessary to use a representation of the generators of $S$ and $T$ in $S_3$ that satisfies $S^2=T^2=(ST)^2=\mathbf{ 1}$. The representation to use is the following
\begin{equation}
\rho(S) = \frac{1}{2} \left( 
\begin{array}{cc}
1 & -\sqrt 3 \\
-\sqrt 3 & -1
\end{array}\right), \qquad 
\rho(T) =  \left( 
\begin{array}{cc}
-1 & 0\\
0 & 1
\end{array}\right),
\end{equation}
where it can be seen that they are also the generators of $S_3$. In this way, for a doublet defined as $Y(\tau)=(Y_1(\tau), Y_2(\tau))^T$ the previous expressions will be as
\begin{align}
&a+b+c=0 \label{pararela2} \notag \\
&Y(S\tau)=Y(-1/\tau)=\tau^2\rho(S) Y(\tau)\\
&Y(T\tau)=Y(\tau+1)= \rho(T) Y(\tau) \notag
\end{align}
The last two equations generate a system of equations relating the parameters $\alpha,\beta$ and $\gamma$. Under $T$, the first component can be rewritten as
\begin{align}
\frac{d}{d\tau}( \log\eta\left( \frac{\tau}{2}\right)[b+a]+\log\eta\left(\frac{\tau+1}{2}\right)[a+b]+\log\eta\left( 2\tau\right)[2c] )=0,
\end{align}
which leads to the relationship $a=-b$ and $c=0$.
Substituting the above relationships, we see that $Y_1(\tau)$ can be written as
\begin{align}
&Y_1(\tau)=\frac{d}{d\tau}\left( a\log\eta\left( \frac{\tau}{2}\right)-a\log\eta\left(\frac{\tau+1}{2}\right) \right)\notag \\
&=C_1 Y(1,-1,0| \tau),
\end{align}
where $C_1=a$.
The other component is calculated with a similar procedure, using (\ref{pararela2}). Thus, the doublet of modular forms of $S_3$ is
\begin{align}
Y_1(\tau)=C_1 Y(1,-1,0| \tau) \hspace{1cm} Y_2(\tau)=C_2 Y(1,1,-2| \tau)
\end{align}
The transformation under $S$ relates the constants $C_1$ and $C_2$ therefore, we obtain the following relationship
\begin{align}
C_1=\sqrt{3}C_2.
\end{align}
The modular forms of weight 2 for $S_3$, with $C_2=i/2\pi$, are
\begin{align} 
&Y_1(\tau) = \frac{\sqrt{3}i}{4\pi}\left( \frac{\eta'(\tau/2)}{\eta(\tau/2)}  -\frac{\eta'((\tau +1)/2)}{\eta((\tau+1)/2)}   \right),\label{formass3a-1}
    \\
&Y_2(\tau) = \frac{i}{4\pi}\left( \frac{\eta'(\tau/2)}{\eta(\tau/2)}  +\frac{\eta'((\tau +1)/2)}{\eta((\tau+1)/2)} - \frac{8\eta'(2\tau)}{\eta(2\tau)} \right),
\label{formass3a}
\end{align}
where the prime indicates the derivative with respect to $\tau$. Figure \ref{Y1Y2} show the real and imaginary part of the modular forms $Y_1(\tau)$ and $Y_2(\tau)$, respectively. Depending on the assignment that is made, the modular forms of weight four will be useful and are obtained from the tensor product of the doublet of modular forms of weight two \cite{kobayashi2019finite}, therefore
\begin{align}
\left(
\begin{array}{c}
Y_1 \\
Y_2
\end{array}\right)
\otimes
\left(
\begin{array}{c}
Y_1 \\
Y_2
\end{array}\right)
=
Y_s^{(4)}+
\left(
\begin{array}{c}
Y_1^{(4)} \\
Y_2^{(4)}
\end{array}\right),
\end{align}
where the antisymmetric singlet vanishes. Furthermore, it has been defined
\begin{align}
 Y_s^{(4)}=Y_1^2+Y_2^2 \hspace{0.7cm} Y_1^{(4)}=2Y_1Y_2 \hspace{0.7cm} Y_2^{(4)}=Y_1^2-Y_2^2. 
\end{align}
through the equation \ref{pararela2} for $S$ in $\tau=i$, we obtain 
\begin{align}
Y_2(i)-\sqrt{3}Y_1(i)=0 \label{relacionformas}.
\end{align}
\section{Complete Lagrangian}\label{complagrangian}
\begin{equation} \begin{aligned} \mathcal{L}_y^{(u)} ={}&C_1\Big\{ [\alpha+\gamma]\big( \overline{Q}_1u_{1R}\tilde{H}_1Y_1^{(4)} +\overline{Q}_2u_{2R}\tilde{H}_2Y_2^{(4)} \big)\\ &+[\alpha-\gamma]\big( \overline{Q}_1u_{1R}\tilde{H}_2Y_2^{(4)} +\overline{Q}_2u_{2R}\tilde{H}_1Y_1^{(4)} \big)\\ &+[\beta+\gamma]\big( \overline{Q}_1u_{2R}\tilde{H}_2Y_1^{(4)} +\overline{Q}_2u_{1R}\tilde{H}_1Y_2^{(4)} \big)\\ &+[\gamma-\beta]\big( \overline{Q}_1u_{2R}\tilde{H}_1Y_2^{(4)} +\overline{Q}_2u_{1R}\tilde{H}_2Y_1^{(4)} \big)\Big\}\\ &+C_2\big( \overline{Q}_1u_{2R}\tilde{H}_1Y_s^{(4)} +\overline{Q}_2u_{1R}\tilde{H}_1Y_s^{(4)}\\ &\qquad +\overline{Q}_1u_{1R}\tilde{H}_2Y_s^{(4)} -\overline{Q}_2u_{2R}\tilde{H}_2Y_s^{(4)} \big)\\ &+C_3\big( \overline{Q}_1u_{2R}\tilde{H}_sY_1^{(4)} +\overline{Q}_2u_{1R}\tilde{H}_sY_1^{(4)}\\ &\qquad +\overline{Q}_1u_{1R}\tilde{H}_sY_2^{(4)} -\overline{Q}_2u_{2R}\tilde{H}_sY_2^{(4)} \big)\\ &+C_4\big( \overline{Q}_1u_{1R}\tilde{H}_sY_s^{(4)} +\overline{Q}_2u_{2R}\tilde{H}_sY_s^{(4)} \big)\\ &+C_5\big( \overline{Q}_1u_{3R}\tilde{H}_2Y_1^{(2)} +\overline{Q}_2u_{3R}\tilde{H}_1Y_1^{(2)}\\ &\qquad +\overline{Q}_1u_{3R}\tilde{H}_1Y_2^{(2)} -\overline{Q}_2u_{3R}\tilde{H}_2Y_2^{(2)} \big)\\ &+C_6\big( \overline{Q}_1u_{3R}\tilde{H}_sY_1^{(2)} +\overline{Q}_2u_{3R}\tilde{H}_sY_2^{(2)} \big)\\ &+C_7\big( \overline{Q}_3u_{1R}\tilde{H}_2Y_1^{(2)} +\overline{Q}_3u_{2R}\tilde{H}_1Y_1^{(2)}\\ &\qquad +\overline{Q}_3u_{1R}\tilde{H}_1Y_2^{(2)} -\overline{Q}_3u_{2R}\tilde{H}_2Y_2^{(2)} \big)\\ &+C_8\big( \overline{Q}_3u_{1R}\tilde{H}_sY_1^{(2)} +\overline{Q}_3u_{2R}\tilde{H}_sY_2^{(2)} \big)\\ &+C_9\overline{Q}_3u_{3R}\tilde{H}_s +\text{h.c.} \end{aligned} \end{equation}

\end{document}